\documentclass[twocolumn]{pasj02} 
\usepackage[switch,mathlines]{lineno} 
\usepackage{natbib} 
\usepackage{bm}

\jyear{2025}
\Received{}
\Accepted{}

\begin{document} 

\title{In-orbit Performance of the Soft X-ray Imaging Telescope  Xtend aboard XRISM}

\author{
Hiroyuki \textsc{Uchida},\altaffilmark{1}\altemailmark\orcid{0000-0003-1518-2188}\email{uchida@cr.scphys.kyoto-u.ac.jp} 
Koji  \textsc{Mori},\altaffilmark{2}
Hiroshi  \textsc{Tomida},\altaffilmark{3}
Hiroshi  \textsc{Nakajima},\altaffilmark{4}
Hirofumi \textsc{Noda},\altaffilmark{5}
Takaaki  \textsc{Tanaka},\altaffilmark{6}
Hiroshi  \textsc{Murakami},\altaffilmark{7}
Hiromasa  \textsc{Suzuki},\altaffilmark{3}
Shogo Benjamin  \textsc{Kobayashi},\altaffilmark{8}
Tomokage   \textsc{Yoneyama},\altaffilmark{9}
Kouichi   \textsc{Hagino},\altaffilmark{10}
Kumiko Kawabata  \textsc{Nobukawa},\altaffilmark{11}
Hideki   \textsc{Uchiyama},\altaffilmark{12}
Masayoshi   \textsc{Nobukawa},\altaffilmark{13}
Hironori   \textsc{Matsumoto},\altaffilmark{14}
Takeshi Go   \textsc{Tsuru},\altaffilmark{1}
 Makoto   \textsc{Yamauchi},\altaffilmark{2}
 Isamu   \textsc{Hatsukade},\altaffilmark{2}
 Hirokazu   \textsc{Odaka},\altaffilmark{14}
 Takayoshi   \textsc{Kohmura},\altaffilmark{15}
  Kazutaka   \textsc{Yamaoka},\altaffilmark{16}
 Tessei   \textsc{Yoshida},\altaffilmark{3}
 Yoshiaki   \textsc{Kanemaru},\altaffilmark{3}
 Daiki   \textsc{Ishi},\altaffilmark{3}
 Tadayasu   \textsc{Dotani},\altaffilmark{3}
 Masanobu   \textsc{Ozaki},\altaffilmark{17}
 Hiroshi   \textsc{Tsunemi},\altaffilmark{14}
  Keitaro   \textsc{Miyazaki},\altaffilmark{2}
 Kohei   \textsc{Kusunoki},\altaffilmark{2}
 Yoshinori   \textsc{Otsuka},\altaffilmark{2}
 Haruhiko   \textsc{Yokosu},\altaffilmark{2}
 Wakana   \textsc{Yonemaru},\altaffilmark{2}
 Kazuhiro   \textsc{Ichikawa},\altaffilmark{2}
 Hanako   \textsc{Nakano},\altaffilmark{2}
 Reo   \textsc{Takemoto},\altaffilmark{2}
 Tsukasa   \textsc{Matsushima},\altaffilmark{2}
 Reika   \textsc{Urase},\altaffilmark{2}
 Jun   \textsc{Kurashima},\altaffilmark{2}
 Kotomi   \textsc{Fuchi},\altaffilmark{2}
 Kaito   \textsc{Hayakawa},\altaffilmark{4}
 Masahiro   \textsc{Fukuda},\altaffilmark{4}
  Shun   \textsc{Inoue},\altaffilmark{1}
 Yuma   \textsc{Aoki},\altaffilmark{11}
 Kouta   \textsc{Takayama},\altaffilmark{11}
 Takashi   \textsc{Sako},\altaffilmark{13}
 Marina   \textsc{Yoshimoto},\altaffilmark{14}
 Kohei   \textsc{Shima},\altaffilmark{14}
 Mayu   \textsc{Higuchi},\altaffilmark{15}
 Kaito   \textsc{Ninoyu},\altaffilmark{15}
 Daiki   \textsc{Aoki},\altaffilmark{15}
 Shun   \textsc{Tsunomachi},\altaffilmark{15}
 Takashi   \textsc{Okajima},\altaffilmark{18}
 Manabu   \textsc{Ishida},\altaffilmark{3}
 Yoshitomo   \textsc{Maeda},\altaffilmark{3}
 Takayuki   \textsc{Hayashi},\altaffilmark{18,19,20}
 Keisuke   \textsc{Tamura},\altaffilmark{18,19,20}
 Rozenn   \textsc{Boissay-Malaquin},\altaffilmark{18,19,20}
 Toshiki   \textsc{Sato},\altaffilmark{21}
 Mai   \textsc{Takeo},\altaffilmark{22}
 Asca  \textsc{Miyamoto},\altaffilmark{23}
 Gakuto \textsc{Matsumoto},\altaffilmark{23}
 Megan E. \textsc{Eckart},\altaffilmark{24}
 Natalie \textsc{Hell},\altaffilmark{24}
 Maurice A. \textsc{Leutenegger},\altaffilmark{18}
 and
 Kiyoshi   \textsc{Hayashida},\altaffilmark{14}}
\altaffiltext{1}{Department of Physics, Graduate School of Science, Kyoto University, Kitashirakawa Oiwake-cho, Sakyo-ku, Kyoto 606-8502, Japan}
\altaffiltext{2}{Faculty of Engineering, University of Miyazaki, 1-1 Gakuen Kibanadai Nishi, Miyazaki, Miyazaki 889-2192, Japan}
\altaffiltext{3}{Japan Aerospace Exploration Agency (JAXA), Institute of Space and Astronautical Science (ISAS), 3-1-1 Yoshino-dai, Chuo-ku, Sagamihara, Kanagawa 252-5210, Japan}
\altaffiltext{4}{College of Science and Engineering, Kanto Gakuin University, Kanazawa-ku, Yokohama, Kanagawa 236-8501, Japan}
\altaffiltext{5}{Astronomical Institute, Tohoku University, 6-3 Aramakiazaaoba, Aoba-ku, Sendai, Miyagi 980-8578, Japan}
\altaffiltext{6}{Department of Physics, Konan University, 8-9-1 Okamoto, Higashinada, Kobe, Hyogo 658-8501}
\altaffiltext{7}{Faculty of Informatics, Tohoku Gakuin University, 3-1 Shimizukoji, Wakabayashi-ku, Sendai, Miyagi 984-8588}
\altaffiltext{8}{Department of Physics, Faculty of Science, Tokyo University of Science, Kagurazaka, Shinjuku-ku, Tokyo 162-0815, Japan}
\altaffiltext{9}{Faculty of Science and Engineering, Chuo University, 1-13-27 Kasuga, Bunkyo-ku, Tokyo 112-8551, Japan}
\altaffiltext{10}{Department of Physics, University of Tokyo, 7-3-1 Hongo, Bunkyo-ku, Tokyo 113-0033, Japan}
\altaffiltext{11}{Department of Physics, Kindai University, 3-4-1 Kowakae, Higashi-Osaka, Osaka 577-8502, Japan}
\altaffiltext{12}{Science Education, Faculty of Education, Shizuoka University, Suruga-ku, Shizuoka, Shizuoka 422-8529, Japan}
\altaffiltext{13}{Faculty of Education, Nara University of Education, Nara, Nara 630-8528, Japan}
\altaffiltext{14}{Department of Earth and Space Science, Osaka University, 1-1 Machikaneyama-cho, Toyonaka, Osaka 560-0043, Japan}
\altaffiltext{15}{Department of Physics, Faculty of Science and Technology, Tokyo University of Science, 2641 Yamazaki, Noda, Chiba 270-8510, Japan}
\altaffiltext{16}{Department of Physics, Nagoya University, Chikusa-ku, Nagoya, Aichi 464-8602, Japan}
\altaffiltext{17}{Advanced Technology Center, National Astronomical Observatory of Japan, Mitaka, Tokyo 181-8588, Japan}
\altaffiltext{18}{X-ray Astrophysics Laboratory, NASA / Goddard Space Flight Center (GSFC), Greenbelt, Maryland, USA}
\altaffiltext{19}{Center for Space Science and Technology, University of Maryland, Baltimore County (UMBC), Baltimore, Maryland, USA}
\altaffiltext{20}{Center for Research and Exploration in Space Science and Technology, NASA/GSFC (CRESST II), Greenbelt, Maryland, USA}
\altaffiltext{21}{Department of Physics, Meiji University, 1-1-1, Higashi Mita, Tama-ku, 214-8571, Kanagawa, Japan}
\altaffiltext{22}{Department of Physics, University of Toyama, 3190 Gofuku, Toyama-shi, Toyama 930-8555, Japan}
\altaffiltext{23}{Department of Physics, Tokyo Metropolitan University, 1-1 Minami-Osawa, Hachioji, Tokyo 192-0397, Japan}
\altaffiltext{24}{Lawrence Livermore National Laboratory, 7000 East Avenue, Livermore CA 94550, USA}



\maketitle

\begin{abstract}
We present a summary of the in-orbit performance of the soft X-ray imaging telescope Xtend onboard the XRISM mission, based on in-flight observation data, including first-light celestial objects, calibration sources, and results from the cross-calibration campaign with other currently-operating X-ray observatories.
XRISM/Xtend has a large field of view of 38.5$\arcmin \times$38.5$\arcmin$, covering an energy range of 0.4--13~keV, as demonstrated by the first-light observation of the galaxy cluster Abell~2319.
It also features an energy resolution of 170--180~eV at 6~keV, which meets the mission requirement and enables to resolve He-like and H-like Fe K$\alpha$ lines.
Throughout the observation during the performance verification phase, we confirm that two issues identified in  SXI onboard the previous Hitomi mission --- light leakage and crosstalk events --- are addressed and suppressed in the case of Xtend.
A joint cross-calibration observation of the bright quasar 3C273 results in an effective area measured to be $\sim420$~cm$^{2}$@1.5keV and $\sim310$~cm$^{2}$@6.0keV, which matches values obtained in ground tests.
We also continuously monitor the health of Xtend by analyzing overclocking data, calibration source spectra, and day-Earth observations: the readout noise is stable and low, and contamination is negligible even one year after launch.
A low background level compared to other major X-ray instruments onboard satellites, combined with the largest grasp ($\Omega_{\rm eff}\sim60~{\rm cm^2~degree^2}$) of Xtend, will not only support Resolve analysis, but also enable significant scientific results on its own. 
This includes near future follow-up observations and transient searches in the context of time-domain and multi-messenger astrophysics.
\end{abstract}


\footnotetext[$\dag$]{Corresponding authors: Hiroyuki Uchida, Hiromasa Suzuki, Shogo Benjamin Kobayashi, Tomokage Yoneyama, Kumiko Kawabata Nobukawa, Hideki Uchiyama, Yoshiaki Kanemaru, and Kouichi Hagino}

\section{Introduction}
The X-ray imaging technique has been crucial for astrophysics \citep{Giacconi1960}, ranging from studies of high-energy extended sources such as  clusters of galaxies and supernova remnants (SNRs) to the identification of transient point sources, ever since the early rocket experiments \citep{Rappaport1979} and satellite missions \citep{Giacconi1979} utilized Wolter-I-type mirrors \citep{Wolter1952}.
In the context of X-ray astronomy, charge-coupled devices (CCDs) have been widely used since their first deployment on the Advanced Satellite for Cosmology and Astrophysics \cite[ASCA;][]{Tanaka1994}; CCDs are still integral to state-of-the-art missions, such as extended ROentgen Survey with an Imaging Telescope Array \citep[eROSITA;][]{Predehl2021} and Space-based multiband astronomical Variable Objects Monitor \citep[SVOM;][]{Atteia2021}.

The X-ray Imaging and Spectroscopy Mission (XRISM) was officially developed starting in 2018 \citep{Tashiro2018} as a continuation of the previous ASTRO-H/Hitomi mission \citep{Takahashi2018} and was successfully launched from the Tanegashima Space Center in Japan  \citep{Tashiro2024} on September 6, 2023, 23:42 (Universal time coordinated).
While the Hitomi mission was prematurely terminated due to an accident with the attitude control system, resulting in a mission lifetime of only about one month, XRISM builds on its technological and scientific heritage in order to recover the expected scientific output that Hitomi had pioneered.
In the early  design specification phase of XRISM (previously referred to as the X-ray Astronomy Recovery Mission, XARM), an X-ray micro-calorimeter and an aligned X-ray CCD camera were selected to be installed on the satellite for high-resolution X-ray spectroscopy and wide-field imaging, respectively: these instruments are named Resolve and Xtend.

Xtend consists of the Wolter-I-type X-ray Mirror Assembly \citep[XMA;][]{Boissay-Malaquin2022, Tamura2022, Hayashi2022} and Soft X-ray Imager \citep[SXI;][]{Hayashida2018}, which essentially builds upon the previous design for Hitomi \citep{Tanaka2018}, with several improvements in order to address issues identified from the observational data of Hitomi \citep{Nakajima2018}.
Since the detailed specifications and ground calibration results for XRISM have been reported during its design phase \citep{Nakajima2020, Mori2022, Mori2024} and summarized in a separate paper \citep{Noda2025}, we focus here on the in-orbit calibration and performance of the XMA$+$SXI system, an integrated mirror and camera assembly, i.e., Xtend.

In this paper, the technical specifications of Xtend, along with its design differences from Hitomi SXI, are summarized in Section~\ref{sec:system}.
The in-orbit performance with Xtend and their calibration results based on data obtained during the commissioning and performance verification (PV) phases are described in Sections~3 and 4, respectively, where we also explain how  the issues observed in the Hitomi data were addressed for XRISM.
Note that the procedure for the initial activation of SXI and its operation policy have been reported by \citet{Suzuki2024}.
Also, the in-orbit performance of the XMA for Xtend has been described by \citet{Tamura2024}.
We therefore omit a detailed discussion of these topics in this paper.
A summary of the in-orbit performance of Xtend and its future prospects as a wide-filed X-ray imaging sensor for astrophysics are presented in Section~5.
The errors quoted in the text and error bars displayed in the figures represent a 1-$\sigma$ confidence level unless otherwise noted.

\section{The Soft X-ray Imaging Telescope, Xtend}\label{sec:system}
\begin{figure}[t]
 \begin{center}
  \includegraphics[angle=0,width=8.2cm]{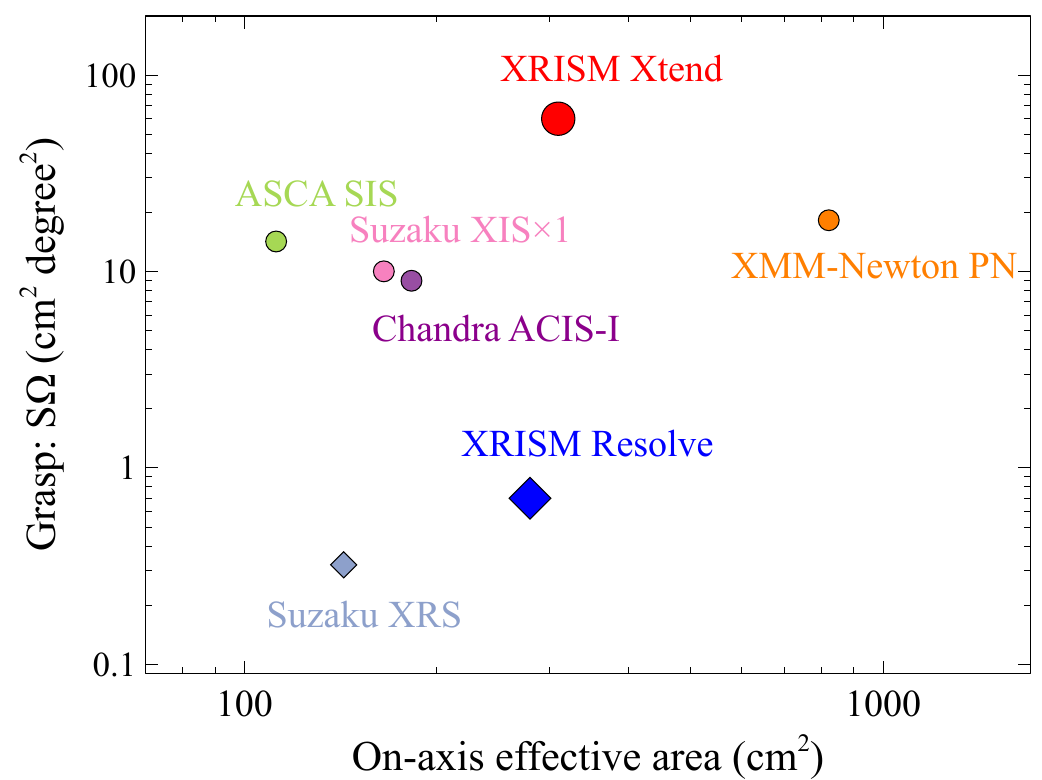} 
 \end{center}
\caption{Comparison between the grasp (${\rm S}\Omega$) and the on-axis effective areas of instruments onboard X-ray satellites. Those of XRISM Xtend and Resolve are colored in red and blue, respectively. The circle marks represent CCD sensors (ASCA SIS, Chandra ACIS-I, Suzaku XIS, XMM-Newton PN, and XRISM Xtend), while the diamond marks represent microcalorimeters (Suzaku XRS and XRISM Resolve). The values, except for those of XRISM, are based on the results reported by \citet{Nakajima2018}.
{Alt text: Graph depicting the relationship between the on-axis effective area and grasp.} 
}\label{fig:grasp}
\end{figure}

As noted in the  introduction above, Xtend consists of the Wolter-I-type mirror (XMA) and the X-ray CCD camera (SXI), whose designs are primarily based on those of Hitomi, with several improvements.
Therefore, in this section, we mainly focus on providing a detailed explanation of the differences between the two.
The main design concepts of Xtend are to cover a wider field of view (FoV) than that of Resolve ($3.1\arcmin\times3.1\arcmin$) in order to enable the search for bright point sources that may contaminate target source emissions \citep[e.g.,][]{XRISM2025}, to estimate extended diffuse backgrounds such as Galactic ridge X-ray emission beyond Resolve's FoV, and improve the total photon statistics.
Xtend itself is, however, also useful for wide-field imaging due to its large grasp (${\rm S}\Omega$), as explained below (figure~\ref{fig:grasp}), and is also important for detecting soft X-rays below 2~keV, particularly in the situation  where the Gate Valve (GV; X-ray aperture door) of Resolve remains closed.

\begin{table*}
  \tbl{Summary of Xtend Performance Specifications Measured/Confirmed from In-orbit Data}{%
  \begin{tabular}{lc}
      \hline
      \hline
     Performance Specifications&XRISM Xtend \\ 
      \hline
      Field of View & $38.5\arcmin\times38.5\arcmin$\footnotemark[$\dagger$]  \\      
       Field of View ($\Omega$)& $0.4~{\rm degree^2}$  \\      
     On-axis Effective Area ($A_{\rm eff}$) & $\sim420$~cm$^{2}$@1.5~keV, $\sim310$~cm$^{2}$@6.0~keV  \\
      Grasp (${\rm S}\Omega$) & $\sim60~{\rm cm^2~degree^2}$@6.0~keV   \\
      Angular Resolution  & $1.4\arcmin$ (HPD),  $7.2\arcsec$ (FWHM) \\
       Energy Range & 0.4--13~keV  \\
      Energy Resolution (calibration source regions) & 170--180~eV@5.9~keV (FWHM)\\
       Energy Resolution (on-axis region) &  $\sim170$~eV@6.4~keV (FWHM)\\
X-ray Background Level & 1.5$\times$10$^{-7}$~counts~s$^{-1}$~keV$^{-1}$~arcmin$^{-2}$~cm$^{-2}$@6.0~keV\\
               \hline
    \end{tabular}}\label{tab:summary}
\begin{tabnote}
\footnotemark[$\dagger$] $18.9\arcmin\times18.9\arcmin$ for a single CCD.  \\ 
\end{tabnote}
\end{table*}

\subsection{X-ray Mirror Assembly (XMA)}
The base design of the XMA originates from mirror assemblies  equipped on  the soft X-ray telescope \citep[SXT;][]{Okajima2016} aboard Hitomi: a conically-approximated Wolter-I-type optic with a gold layer  deposited on a glass mandrel and  replicated onto the thin aluminum substrate.
One of the changes for XRISM is that a glass sheet wrapped around an aluminum mandrel was used for replication instead of a Pyrex glass tube \citep{Okajima2020}.
Based on the ground measurement for the SXT, the half-power diameter (HPD) and  the full width at half maximum (FWHM) were estimated to be $1.4\arcmin$ and  $7.2\arcsec$, respectively (private communication with the Xtend XMA team).

\subsection{Soft Xray Imager (SXI)}\label{sec:sxi}
The SXI system consists of the CCD camera (SXI-S), the pixel processing electronics (SXI-PE), the digital electronics (SXI-DE), and the cooler driver (SXI-CD).
The contamination blocking filter (CBF) is equipped at the top of the SXI-S camera body.
The SXI-S includes the CCD sensors, a single-stage Stirling cooler (SXI-S-1ST), analog electronics for driving the sensors (SXI-S-FE), and video boards for processing the output signals.
We applied four P-channel back-illuminated (BI) CCDs manufactured by Hamamatsu Photonics K. K. for  the X-ray detector (hereafter, CCD1--4).
To suppress the charge transfer inefficiency (CTI), a notch structure \citep{Kanemaru2019} was incorporated into the charge transfer path unlike the design of Hitomi \citep[see also,][]{Noda2025}.
Each CCD has a 200-$\mu$m-thick depletion layer, arranged in a $2\times2$ array on the focal plane of Xtend within the camera body of the SXI-S, with a size of 62~mm$\times$62~mm.
A frame exposure is 3.96~sec in the nominal operation case and signals are moved to the frame store area in 36.864~msec for readout.
To enhance the optical blocking performance, a thick aluminum optical blocking layer (OBL) was applied to the surface of the CCDs \citep[100$+$100-nm OBL;][]{Uchida2020}.

The four CCD chips are cooled by the SXI-S-1ST, which is powered by  the SXI-CD and regulated through a heater controlled by the SXI-DE to maintain a constant temperature ($-110~{\rm\degree C}$ during a normal operation).
Unlike Hitomi, Xtend does not carry two SXI-S-1ST units for redundancy, but instead carries a single unit.
The SXI-PE generates a CCD clock pattern and processes digitized frame data from the video board, applying the User Field Programmable Gate Array (UserFPGA) on the mission I/O (MIO) board.
Each CCD chip has two segments, AB and CD in normal observations, with X-ray events read out through different nodes.
Further data processing is handled by the SXI-DE, which executes commands from the satellite management unit (SMU) and transmits telemetry, including frame and housekeeping (HK) data, to the SMU and the spacecraft data recorder (DR).
The detailed design, specifications, and roles of these SXI components are summarized in a separate paper \citep{Noda2025}.

 \begin{figure*}[t]
 \begin{center}
  \includegraphics[width=8cm]{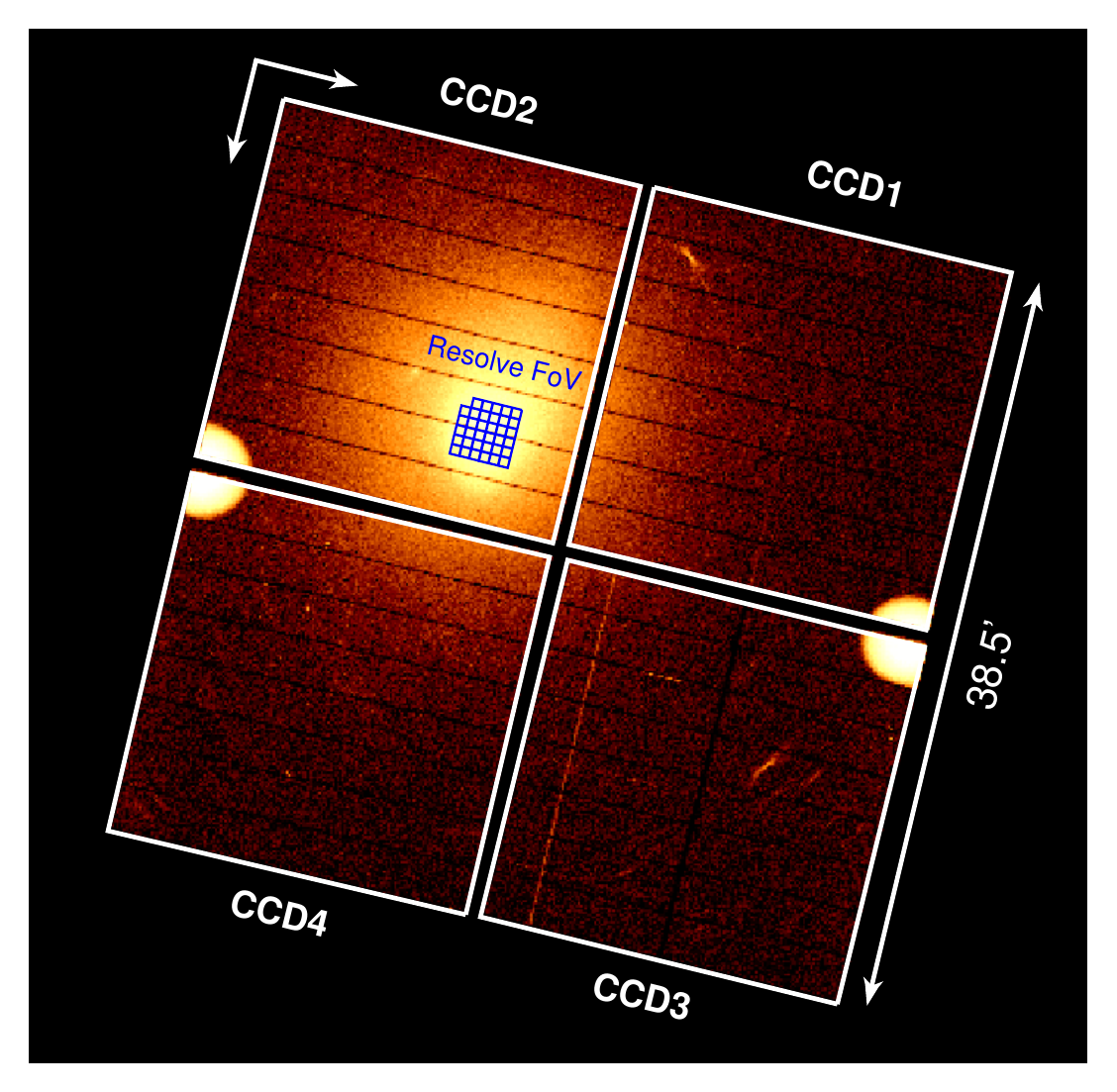} 
    \includegraphics[width=8.8cm]{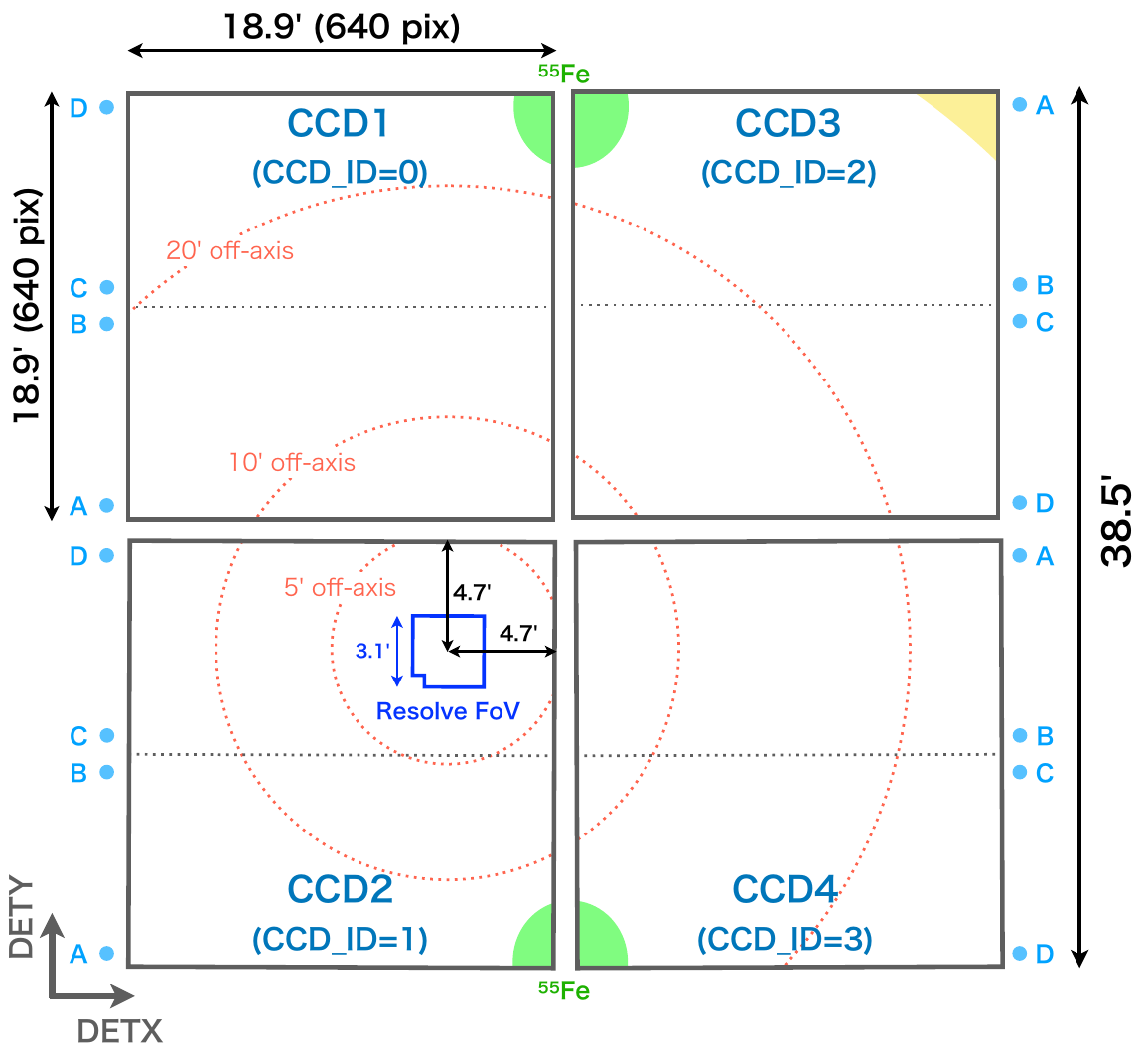} 
 \end{center}
\caption{Left: Full-band (0.4--13~keV) Xtend image of a cluster of galaxies, Abell~2319, obtained in the normal mode, before the vignetting correction. The image is displayed in the equatorial coordinate system. The FoVs of each CCD are outlined with the white squares, and that of Resolve is overlaid with the blue squares. The gap between the segments of CCD3 is eliminated due to a bad column. Right: Schematic layout of the Xtend SXI CCDs as viewed from the back of the focal plane, looking toward the sky. The four CCD chips are arranged in a $2\times2$ format. The names of the CCDs (CCD1--4) and their corresponding identifications used in the XRISM analysis software (CCD\_ID~0--3) are also labeled. Each CCD is divided into two electrically separated segments, with the boundaries indicated by dotted grey lines. The readout nodes are labeled as A, B, C, and D in light blue. The filled green semicircles indicate the approximate areas illuminated by the $^{55}$Fe calibration sources. The area enclosed by the blue line represents the Resolve FoV, and the dotted red circles indicate the off-axis distances. The yellow area at the corner of CCD3 marks the region where X-ray photons are obscured by the camera housing (see text). 
{Alt text: X-ray Image of Abell~2319 and schematic layout of the detector.} 
}\label{fig:abell2319_img}
\end{figure*}

\subsection{Xtend}
The SXI-S is positioned at the focal plane of the XMA, which has a focal length of 5.6~m, providing an FoV of $38.5\arcmin\times38.5\arcmin$ covering an energy range of $0.4$--$13$~keV: $18.9\arcmin\times18.9\arcmin$ (640 logical pixels for each side\footnote{Each CCD has $1280\times1280$ pixels, which are binned into $2\times2$ pixel blocks (i.e., logical pixels) during readout.}) for each CCD.
Thanks to its large FoV  and effective area,  Xtend provides significant collecting power, parameterized by the grasp (at 6~keV)
\begin{eqnarray}
{\rm S}\Omega=\int_{\Omega} A_{\rm eff}(r, \theta)~d\Omega \sim60~{\rm cm^2~degree^2},
\end{eqnarray}
 where $A_{\rm eff}(r, \theta)$ represents the effective area at a position $r$, the radial distance from the on-axis center $r=0$,  an azimuth angle $\theta$, and $\Omega$ is the FoV in degrees.
 As a result, the grasp ${\rm S}\Omega$ of Xtend is  the largest  of any currently available focal plane X-ray detector as summarized in figure~\ref{fig:grasp}, demonstrating its capability to observe extended diffuse sources such as nearby clusters of galaxies, SNRs, and Galactic center plasmas \cite[e.g.,][]{Audard2024}, as well as enabling transient searches in the context of time-domain  astronomy \citep[see also][]{Tsuboi2024}.
Performance specifications of Xtend including the grasp are summarized in Table~\ref{tab:summary}, which are explained in detail in this paper.

\section{Observations}\label{sec:obs}
After the critical operation period for XRISM, the commissioning phase began, during which we initiated the power-on operation for SXI including the health check of the CCDs on October 17, 2023, one month after the launch, and completed it on October 23, 2023.
We confirmed that there was no anomalous charge intrusion \citep[][for more details]{Noda2024} observed in several ground tests of the SXI.
Parameter optimization and adjustment of the CCDs were performed from October 23 until December 3, 2023, and we successfully transitioned to the normal observation phase on December 3, 2023.
The detailed procedures for these initial operations are outlined in our recent paper \citep{Suzuki2024}.
XRISM completed its commissioning period and transitioned to the nominal phase on March 4, 2024. 
Since November 22, 2023, the Xtend CCDs have been operating in event mode, which reads out only pixels that exceed a pre-set threshold and their neighboring pixels, except for status checks during calibration observations.

We aimed XRISM at a neighboring cluster of galaxies, Abell~2319 and a core-collapse SNR in the Large Magellanic Cloud, N132D, for the ``first light'' observations, with exposure times of 92.6~ks and 52.6~ks, respectively.
During the following PV phase, several point sources, including 3C273 (quasar) and PKS~2155$-$304 (blazar), as well as the extended source, the Cygnus Loop (SNR) and the Perseus Cluster, were observed for calibration purposes to mainly estimate the effective area.
Thermal plasmas, such as 1E~0102.2$-$7219 (SNR), were also observed to calibrate the detector response and to estimate the contamination level.
We note that day/night Earth data and observations of blank sky or source free regions are also used to evaluate the optical blocking performance and to measure the background level, including the solar wind charge exchange (SWCX; Ishi et al., \textit{in prep.}, for more details) and non-X-ray background (NXB).
All the event data used hereafter were screened using the XRISM calibration database as part of the initial data processing with \texttt{xapipeline}, provided in version 6.34 of the High Energy Astrophysics Software \citep[HEASoft/ftools;][]{Blackburn1995}.
The following spectral fits were done on the platform of the X-ray SPECtral fitting package (XSPEC) 12.14.1 \citep{Arnaud1996}. 

\subsection{Demonstration of Xtend's Imaging and Spectral Capabilities}\label{sec:demo}
\begin{figure}[t]
 \begin{center}
  \includegraphics[width=7cm]{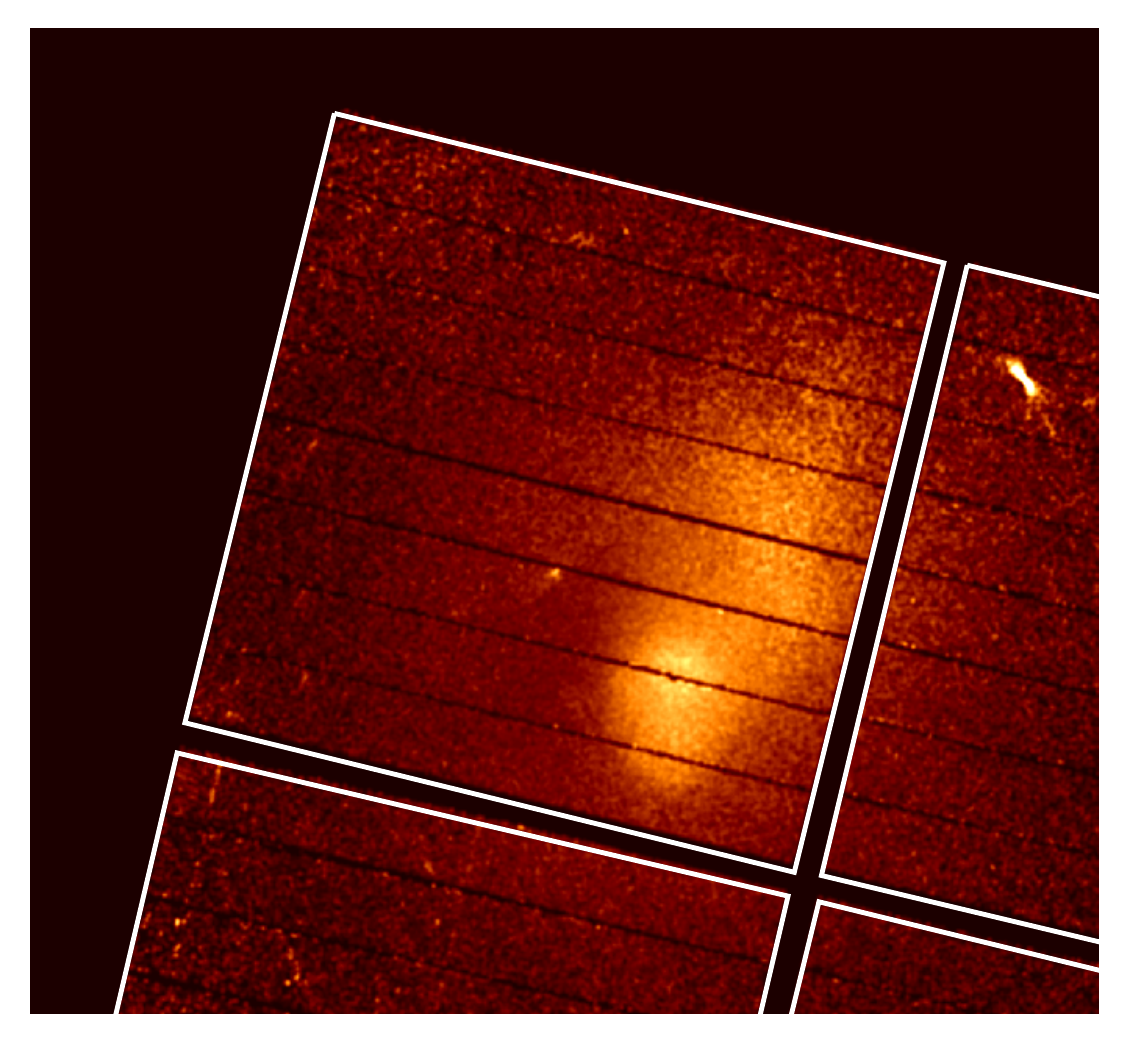} 
 \end{center}
\caption{Vignetting-corrected Xtend image of a cluster of galaxies, Abell~2319. The observation is the same as that shown in figure~\ref{fig:abell2319_img}, but the energy band around the Mn K$\alpha$ line from $^{55}$Fe has been eliminated. 
 {Alt text: X-ray Image of Abell~2319.} 
}\label{fig:abell2319_img_vig}
\end{figure}

Figure~\ref{fig:abell2319_img} displays the Xtend image of Abell~2319 (ObsID: 000103000), in which the extended structure  of the merging galaxy cluster is detected beyond the FoV of Resolve.
Mn~K$\alpha, \beta$ events from the onboard $^{55}$Fe calibration sources are also observed at the edges of the FoV.
A schematic layout of the FoV is shown in the right panel of figure~\ref{fig:abell2319_img}.
Each CCD is slightly rotated counterclockwise when viewed towards the image plane (look-up view) relative to CCD2, with angles of $0.35^\circ$ (CCD1), $0.05^\circ$ (CCD3), and $0.60^\circ$ (CCD4).
We caveat that CCD1--4 correspond to CCD\_ID~0--3 in the XRISM analysis software.
In figure~\ref{fig:abell2319_img}, a few point sources are also detected within the Xtend FoV; their point spread functions become increasingly distorted toward its edges.
The result clearly demonstrates the overwhelmingly large FoV and an improved angular resolution.
As shown in figure~\ref{fig:abell2319_img_vig}, the vignetting-corrected image reveals an arc-like discontinuity and a tail structure extending toward the northwest, which is a signature of a merger of the clusters as reported in the previous observations \citep[e.g.,][]{OHara2004}.

Note that the Xtend FoV contains narrow mechanical gaps of approximately $40\arcsec$ between adjacent chips, and the XRISM nominal (on-axis) point --- defined as the center of Resolve FoV --- is offset by  $\sim4.7\arcmin$ from the sides of CCD2.
A corner of CCD3 is obscured by the camera housing; this obstruction is negligible in the image of Abell~2319 (but is noticeable in figures~\ref{fig:cyg_image} and \ref{fig:dayearth}).
In figure~\ref{fig:abell2319_img}, small gaps at regular intervals along the readout direction, every 80 logical pixels, are also visible.
These gaps are caused by the spaced-row charge injection (CI) operation \citep{Prigozhin2008, Uchiyama2009, Nobukawa2014, Kanemaru2020}.
The CI method mitigates signal charge losses during readout due to lattice defects and improves the CTI of in-orbit photon-counting CCDs \citep{Tomida1997, Townsley2002, Plucinsky2003}.
The CI rows, along with the first preceding row and the first and second trailing rows, which all suffer from charge leakage, are therefore excluded under standard screening criteria.
Although one of the injection rows crosses the Resolve FoV in figure~\ref{fig:abell2319_img}, the injection pattern has been modified since March 10, 2024, to avoid any overlap with the CI rows.
More detailed information on this issue is provided by \citet{Suzuki2024}.

\begin{figure}[t]
 \begin{center}
  \includegraphics[angle=0,width=8.5cm]{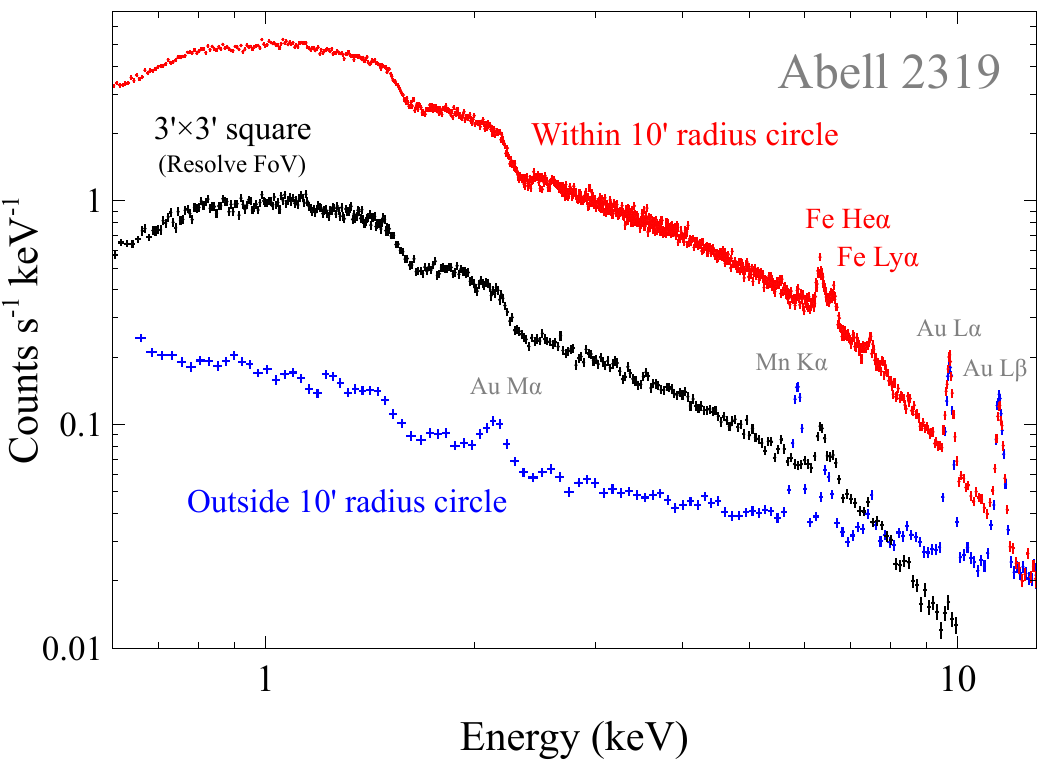} 
 \end{center}
\caption{Xtend spectra of Abell~2319 obtained from different regions within the FoV. The black, red, and blue data points represent the spectra from the Resolve's FoV region ($3\arcmin\times3\arcmin$ square), a $10\arcmin$ radius circle centered at the nominal position, and the region outside the circle, respectively. Fe emission lines originating from Abell~2319 are labeled in red, while NXB fluorescence and $^{55}$Fe isotope lines are labeled in gray.
{Alt text: X-ray spectra.} 
}\label{fig:abell2319_spec}
\end{figure}

Figure~\ref{fig:abell2319_spec} shows example spectra of Abell~2319 extracted from various areas within the Xtend FoV, excluding the calibration source regions.
Thanks to its large effective area, high photon flux is achieved across a broad energy range, including below 2~keV, where soft X-rays are blocked by the Resolve GV.
 Xtend can improve the statisticas in the low-energy band, making it easier to estimate spectral features of continuum emissions.
The spectra of the core of Abell~2319 within a radius of $10\arcmin$  exhibit a high-temperature thermal component \citep{Sugawara2009} characterized by significant emission lines of ionized iron (Fe).
We confirmed that He-like and H-like Fe lines (6.64--6.70~keV and 6.95--6.97~keV in the rest frame, respectively) are clearly separated, demonstrating that Xtend meets the energy resolution requirement of XRISM.
The spectrum obtained from the outer region is dominated by the Cosmic X-ray Background (CXB) and the NXB; particularly Au L$\alpha, \beta$ lines around 10--12~keV originating from the NXB are significant, which will be discussed in detail in Section~\ref{sec:nxb}.
Additionally, only in the spectrum outside the $10\arcmin$ circle, Mn K$\alpha$ and K$\beta$ lines from the calibration sources are present, which are considered out-of-time events, referring to X-rays detected during the charge transfer process. 
These events occur when X-rays are incident during the transfer of charge across the detector.
They are therefore detected downstream of the calibration source regions along the charge transfer direction (horizontal in Figure~\ref{fig:abell2319_img}).

\begin{figure*}[t]
 \begin{center}
  \includegraphics[angle=0,width=8cm]{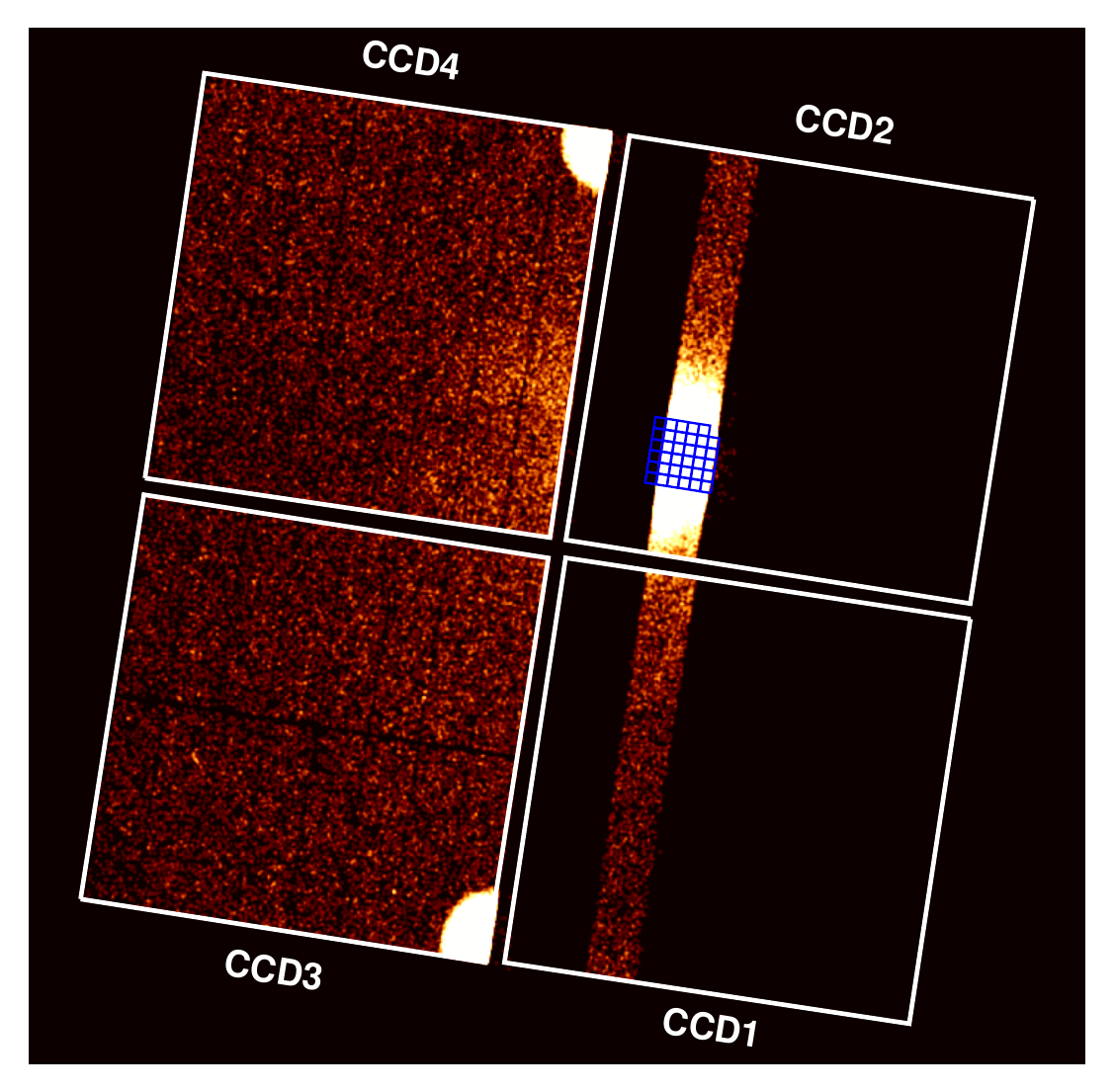} 
   \includegraphics[angle=0,width=9cm]{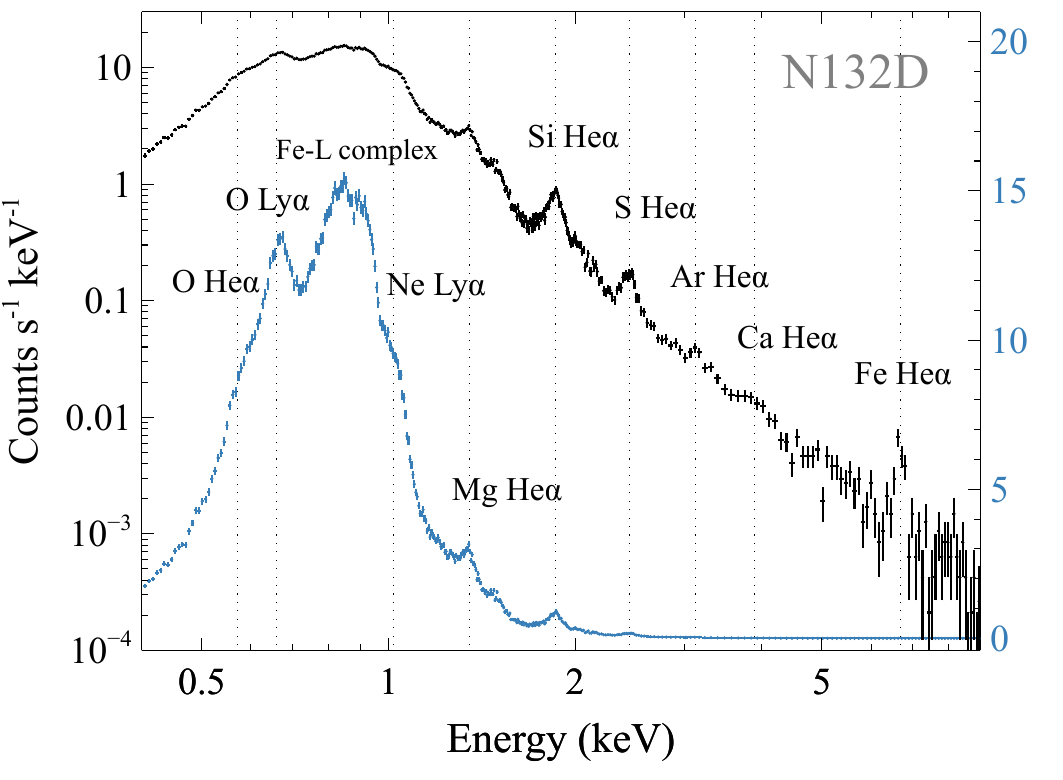} 
\end{center}
\caption{Left: Full-band (0.4--13~keV) Xtend image of N132D and the surrounding background region obtained in the "1/8 Window No Burst" mode. The FoVs of each CCD are outlined by the white squares. The FoV of Resolve is also overlaid on the image with the blue squares. The image is displayed in the equatorial coordinate system. Close-up view of N132D is shown in figure~\ref{fig:N132D_color}.
Right: Xtend spectrum of the entire region of N132D obtained in the "1/8 Window No Burst" mode. The two spectra are derived from the same data but are presented with different y-axis scales: the black points are plotted on a logarithmic scale, while the blue points are plotted on a linear scale. The right y-axis represents the linear scale. The vertical dotted lines indicate the centroid energies of emission lines.
 {Alt text: X-ray Image of N132D and its spectrum.} 
}\label{fig:N132D}
\end{figure*}

\begin{figure*}[t]
 \begin{center}
  \includegraphics[angle=0,width=4.5cm]{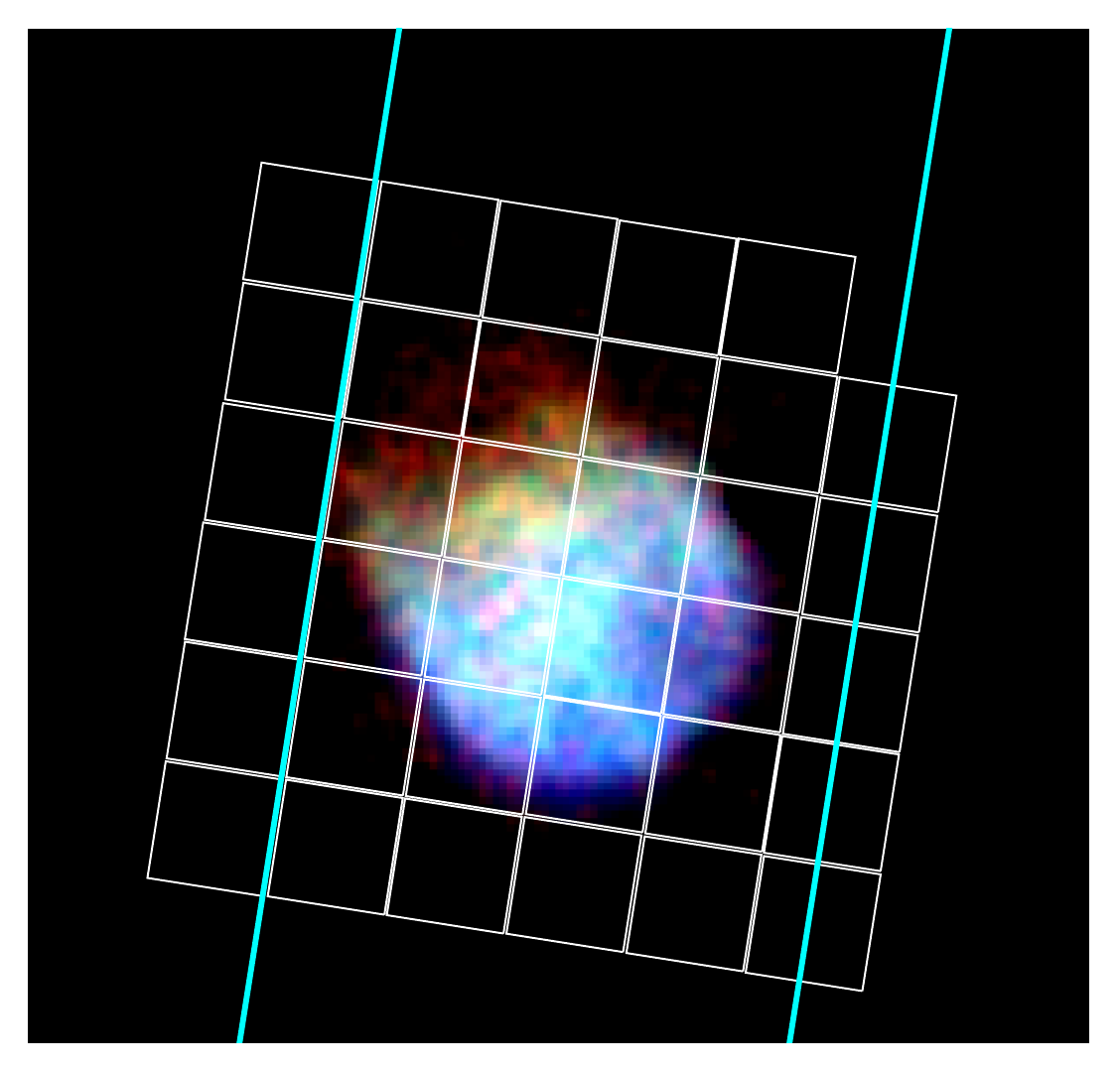} 
    \includegraphics[angle=0,width=4.5cm]{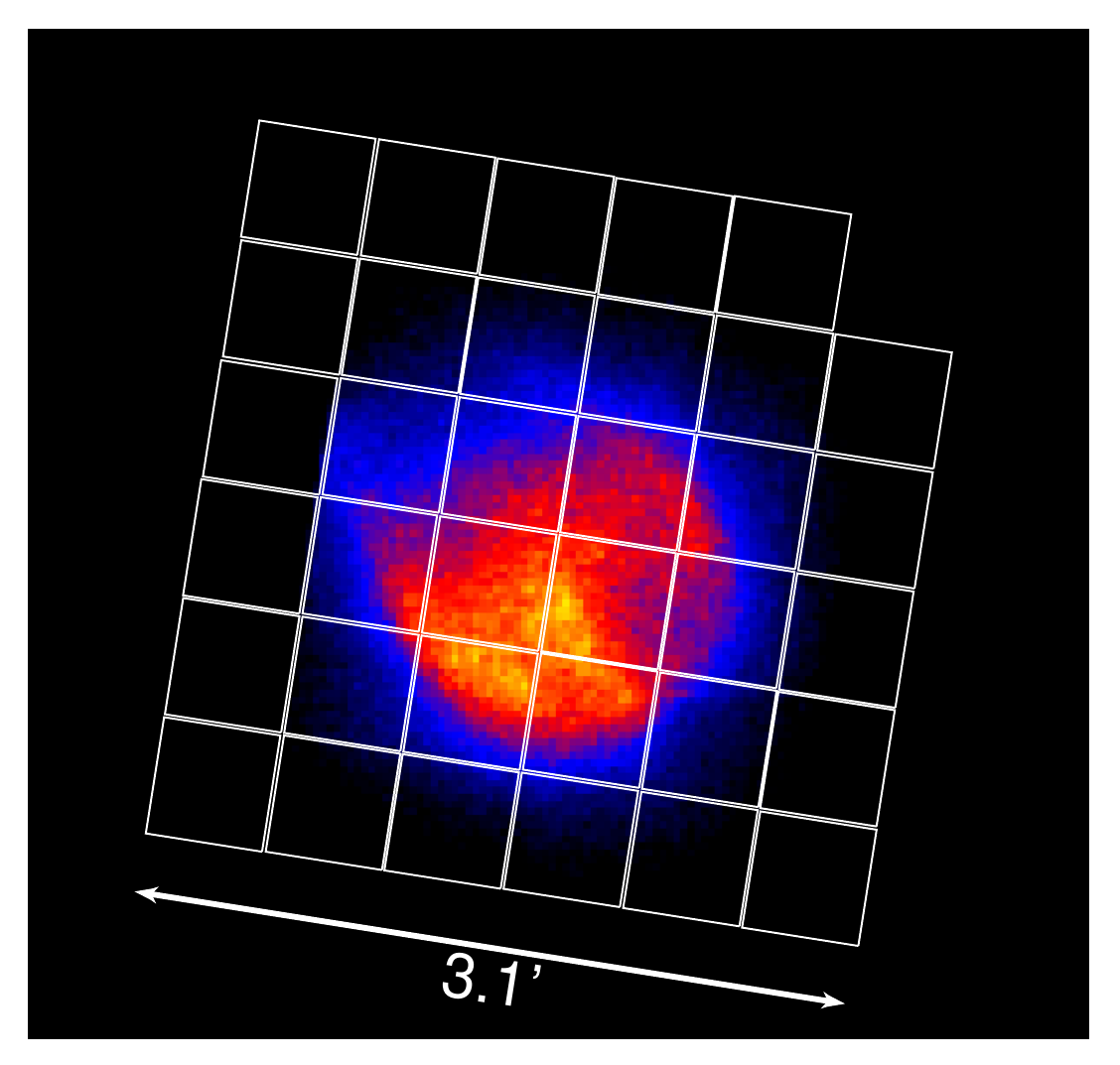} 
  \includegraphics[angle=0,width=4.5cm]{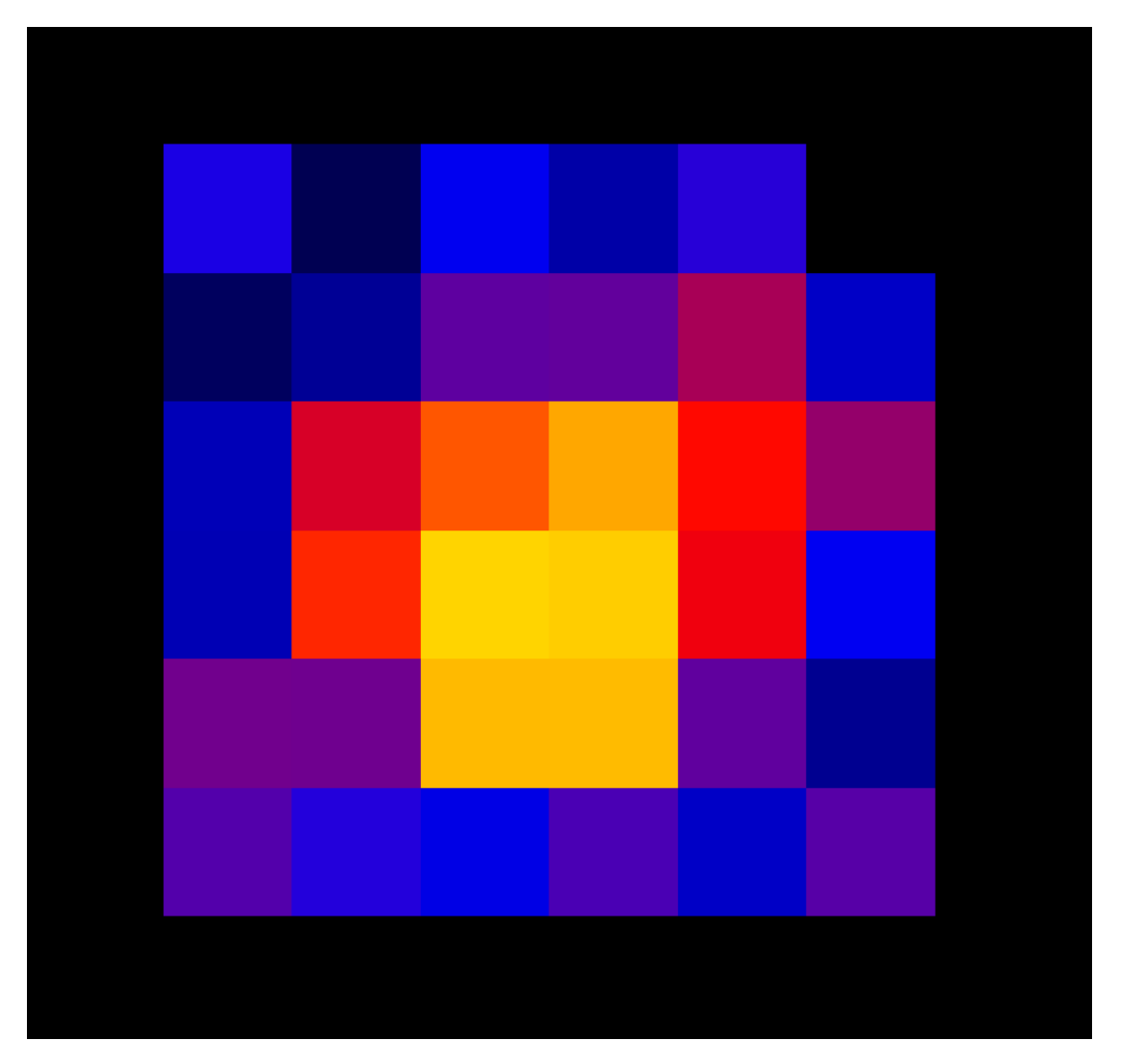} 
 \end{center}
\caption{Close-up view of the Xtend (CCD2)  and Resolve images of N132D. The left, middle, and right panels show the true-color Xtend image, the full-band Xtend image, and the full-band Resolve image, respectively. The Resolve pixel regions are indicated by white squares. In the true-color image, red, green, and blue correspond to 0.4--0.5~keV,  0.5--0.75~keV, blue 0.75--10~keV, respectively.  The FoV of Xtend for CCD2 in the "1/8 Window No Burst" mode is outlined by the cyan lines. The Resolve image is presented in the detector coordinate system.
 {Alt text: X-ray Images of N132D.} 
}\label{fig:N132D_color}
\end{figure*}


There are several optional observation modes available for users to choose from, in addition to the nominal operation mode, referred to as the ``Full Window (No Burst)'' mode \citep[cf.][]{Noda2024}.
Figure~\ref{fig:N132D} shows an example observation of N132D (ObsID: 000128000) obtained in the ``1/8 Window No Burst'' mode, which reads only 80 CCD rows (1/8 of the full logical pixels) per exposure, thereby reducing the readout time to 1/8 and refining the time resolution of Xtend.
This window mode, with or without a burst option (which collects events during a fraction of the nominal exposure time), reduces the likelihood of pile-up events from bright sources\footnote{The pile-up limit for the Full Window mode is approximately 3~mCrab. Detailed information, including a pileup simulator for Xtend, is provided in separate papers \citep[e.g.,][]{Yoneyama2024}} and is also well suited for astronomical objects that exhibit rapid temporal variability.
As seen in figure~\ref{fig:N132D}, the window and burst modes are applied to CCD1 and CCD2 including the XRISM nominal position.
In the right panel of figure~\ref{fig:N132D}, we show an example spectrum of N132D extracted from the on-axis position.
Since this remnant is in a non-equilibrium ionization state \citep{Suzuki2020} and has a lower temperature compared to Abell~2319, emission lines of ionized elements lighter than Fe are detected with Xtend.

Figure~\ref{fig:N132D_color} displays a close-up view of the image, in which a break of the shell \citep{Sharda2020} is clearly seen.
By comparing the Xtend image with the Resolve image obtained during the same observation (right panel of figure~\ref{fig:N132D_color}), we confirm that the flux peak corresponds to the brightest region of the remnant, located at the southern center, which suggests an asymmetric ejecta morphology, as previously reported by \citet{Borkowski2007}.
The scientific results of the XRISM first-light observation of N132D have been published in \citet{Audard2024}, where we present photon count images of lines from several elements using Xtend.

\subsection{Issues Identified in Hitomi SXI}
As documented by \citet{Nakajima2018}, Hitomi SXI exhibited several issues that we first identified in in-orbit data, namely light leakage and crosstalk events\footnote{Sometimes the crosstalk event is termed as ``cosmic(-ray) echo event'' as seen in some XRISM documents: The XRISM Data Reduction Guide (https://heasarc.gsfc.nasa.gov/docs/xrism/analysis/abc\_guide/xrism\_abc.html)}, both of which cause pseudo signals.

\subsubsection{Light leakage} 
The cause of the light leakage was the holes provided for  instruments in the base panel of the Hitomi spacecraft.
These holes are sealed in the XRISM design, thereby blocking the main light paths.
XRISM also has applied countermeasures (sealing gaps and reducing the reflectivity of materials) to the spacecraft's vent holes  to mitigate any incoming light from all directions.
We further implemented design modifications to the CCDs to improve optical blocking performance, as summarized by \citet{Uchida2020}.
While there were two different origins of the light leakage of Hitomi SXI \citep[pinholes and end-surface leakage;][]{Uchida2020}, we confirmed that no significant light leak events are observed in any spacecraft attitude from the in-orbit data obtained with XRISM Xtend.
We therefore conclude that this issue has been addressed for XRISM. 
Since the deterioration of optical blocking performance over time could become significant, we will continue monitoring the light leak until the satellite reaches its end of life (EOL).

\subsubsection{Crosstalk} 
Crosstalk events \citep[see figure~6 in][]{Nakajima2018} are, on the other hand, still observed in raw data of Xtend.
This is because this phenomenon originates from capacitive and/or inductive coupling along the signal paths from the CCDs to the electronics, probably the video board, and therefore the identical design of the electronics results in a similar issue. 
When a cosmic ray particle produces a large amount of signal charge in a pixel, a negative signal is induced by the cross talk in the pixel at a corresponding position in the adjacent segment.
If the resultant low pulse height falls below the dark threshold, the dark level is updated to an anomalously low value, causing pseudo events, i.e., crosstalk events, to be generated continuously thereafter.
This phenomenon therefore can be avoided by adjusting the dark threshold so that a low pulse height is not used for updating the dark level.
Since the incidence of crosstalk events was anticipated before the launch, we examined an appropriate value for the dark threshold.
As a result, pseudo-crosstalk events were largely reduced in the case of Xtend particularly above 0.6~keV and we are further developing a tool to effectively remove these events\footnote{For  detailed information and a method, refer to the XRISM Quick-Start Guide (https://heasarc.gsfc.nasa.gov/docs/xrism/analysis/quickstart/index.html).}. 
We  also note that the on-axis segment (CCD2CD), where the Resolve nominal position is located, does not suffer from this phenomenon due to the polarity of the crosstalk signal.

\section{Calibration}\label{sec:cal}

\begin{figure*}[t]
 \begin{center}
  \includegraphics[angle=0,width=12cm]{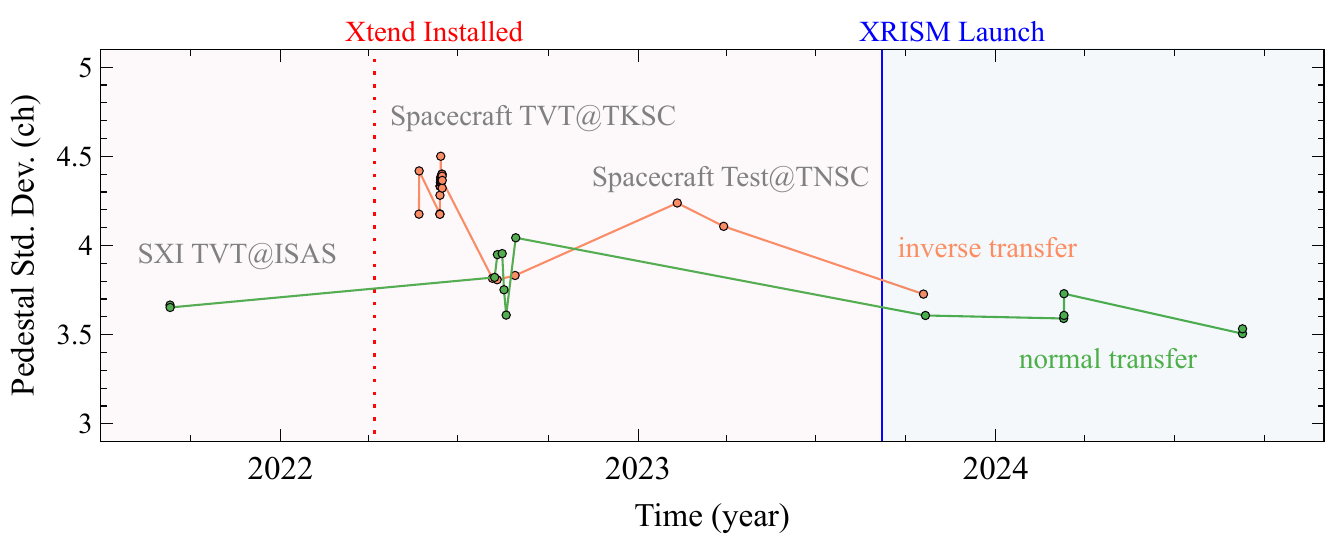} 
 \end{center}
\caption{Long-term trend of the noise level (standard deviations of the pedestal) of the on-axis segment, CCD2CD. The orange and green lines represent values obtained in the inverse and normal transfer modes, respectively. The vertical dotted red and solid blue lines correspond to the Xtend installation on the spacecraft (April 22, 2022) and the XRISM launch date (September 6, 2023), respectively. The hatched areas in light red and light blue indicate the time during ground tests and in orbit, respectively. In the plot, we also labeled the names of several ground tests: the SXI thermal vacuum test (TVT) conducted at the Institute of Space and Astronautical Science (ISAS), the spacecraft TVT at the Tsukuba Space Center (TKSC), and a ground test conducted at the launch suite, the Tanegashima Space Center (TNSC).
{Alt text: Line graph showing a trend of the noise level.} 
}\label{fig:pedestal}
\end{figure*}

\subsection{Readout Noise}
Since the launch of XRISM, we have continuously monitored the readout noise (pedestal) in Xtend frame data as part of routine health checks for the CCDs and the SXI system.
Before cooling the CCDs, we measured the readout noise of the frames by transferring charges in the serial register in the direction opposite to the readout node; this process is referred to as ``inverse transfer mode'' \citep[e.g.,][]{Nakajima2018}.
The noise histogram was obtained in the horizontal overclock (HOC) region:
\begin{eqnarray}
325\leq {\rm READX}\leq 330~\&\&~10\leq {\rm READY}\leq 600,
\end{eqnarray}
where  ${\rm READX}$ and ${\rm READY}$ denote a pixel location along the horizontal and vertical readout directions, respectively.
During the normal operation at a CCD temperature of $-110~{\rm\degree C}$, we obtained a noise histogram in the same HOC region; this process is referred to as ``normal transfer mode''.

Figure~\ref{fig:pedestal} shows the trends in the standard deviation of the pulse height distributions of the pedestal in the on-axis segment, CCD2CD, measured during ground tests and in orbit.
The resultant noise level was 3.7~ch (analog-to-digital unit) before the cooling, and has remained nearly constant, or even improved slightly, at 3.5~ch in the most recent measurement during normal operation.
The long-term trend shows values ranging from 3.7~ch to 4.5~ch in the inverse transfer mode, and 3.5~ch to 4.0~ch in the normal transfer mode. 
Similar trends were also obtained from all other segments.
We note that 1~ch approximately corresponds to 6~eV, resulting in fluctuations in the readout noise ranging from 21~eV to 27~eV.
These results indicate that the noise level of the Xtend CCDs has remained unchanged before and after the launch.

\subsection{Energy Resolution}

\begin{figure}[t]
 \begin{center}
  \includegraphics[angle=0,width=8.5cm]{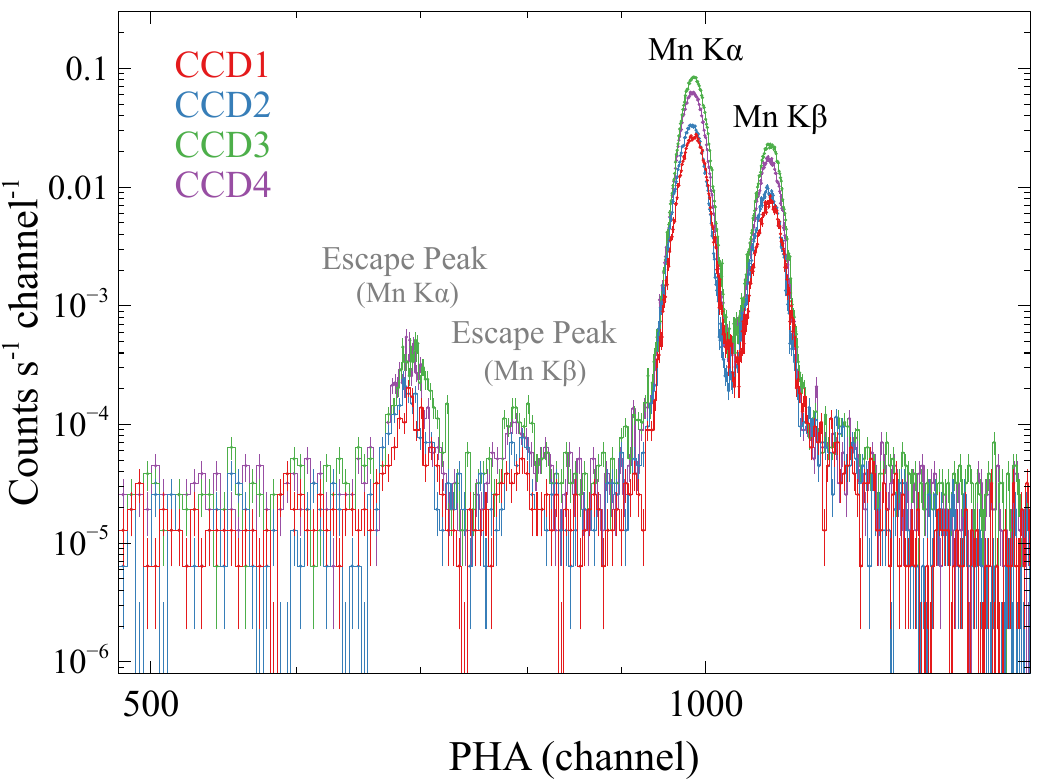} 
 \end{center}
\caption{Example spectra of $^{55}$Fe isotopes (Mn K$\alpha$ and Mn K$\beta$ lines) obtained from the calibration source regions on each CCD of Xtend. The spectra from CCD1, CCD2, CCD3, and CCD4 are shown in red, blue, green, and purple, respectively. The horizontal axis is a pulse height amplitude (PHA).
 {Alt text: X-ray Spectra of $^{55}$Fe isotopes.} 
}\label{fig:55Fe}
\end{figure}

In figure~\ref{fig:55Fe}, we  present example spectra of the calibration source regions on each CCD, illuminated by the $^{55}$Fe isotopes, which were obtained during the PV phase.
Mn K$\alpha$ and Mn K$\beta$ lines at $\sim980$~ch and $\sim1090$~ch, corresponding to 5.9~keV and 6.4 keV, respectively, along with their silicon escape peaks are detected.
The count rate is $\sim1$~counts~sec$^{-1}$ for each CCD at the beginning of the commissioning phase.
Taking into account the half-life of the isotopes, we verified that  the result is in agreement with the extrapolated values from ground-based measurements.
The overall spectral features are consistent with those obtained at the ground experiments of SXI.

\begin{table*}
  \tbl{Energy resolutions of Xtend at the Mn K$\alpha$ line energy, measured in the calibration source regions.}{%
  \begin{tabular}{lcccc}
       \hline
     \hline
      & \multicolumn{4}{c}{Energy Resolution (eV)@5.9~keV (FWHM)}  \\ 
      \cline{2-5}
     &CCD1\footnotemark[$\dag$] & CCD2\footnotemark[$*$]  & CCD3\footnotemark[$*$]  & CCD4\footnotemark[$\dag$]  \\ 
      \hline
      XRSIM in orbit (October 23, 2023) & 184.3$\pm$1.1 & 179.0$\pm$1.0 & 170.7$\pm$0.6 & 173.0$\pm$0.7  \\
      \hline
      TVT (August, 2022; COLD case)  & 186.2$\pm$0.6 & 180.7$\pm$0.5 & 168.7$\pm$0.3 & 171.3$\pm$0.4 \\
      TVT (August, 2022; HOT case)   & 182.8$\pm$1.3 & 176.4$\pm$1.1 & 168.7$\pm$0.7 & 172.6$\pm$0.8 \\
      \hline
    \end{tabular}}\label{tab:55Fe}
\begin{tabnote}
\footnotemark[$*$] segment AB  \\ 
\footnotemark[$\dag$] segment CD
\end{tabnote}
\end{table*}

\begin{figure*}[t]
 \begin{center}
  \includegraphics[angle=0,width=11cm]{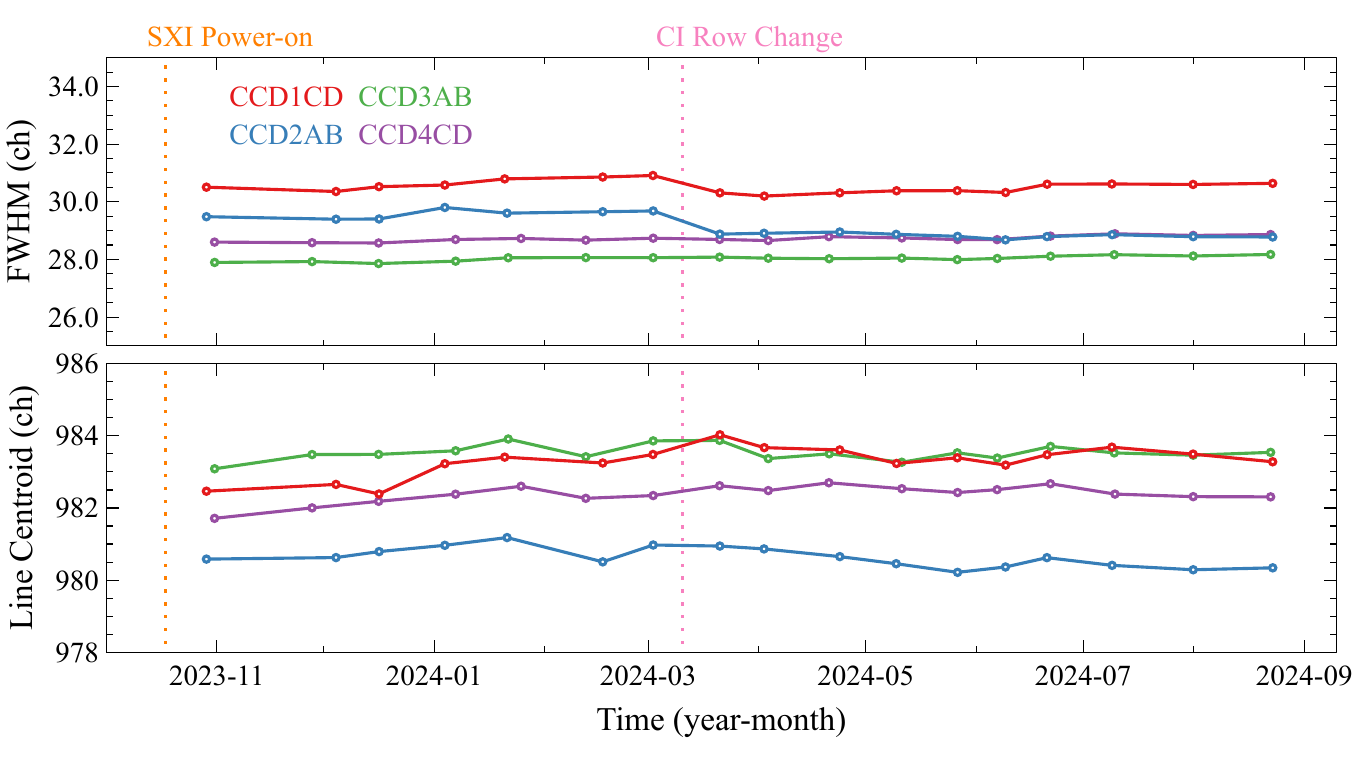} 
 \end{center}
\caption{Long-term trends of the FWHM (top) and the line centroid (bottom) of the Mn K$\alpha$ line in the calibration source regions on each CCD. The vertical dotted orange and pink lines correspond to the time when the SXI system  was turn on (October 17, 2023) and when the CI rows (injection pattern) were changed (March 10, 2024), respectively.
{Alt text: Line graphs showing a the FWHM and the line centroid of the Mn K$\alpha$ line.} 
}\label{fig:55Fe_trend}
\end{figure*}

Since the CCDs are exposed to radiation damage in orbit, the CTI is expected to increase over time.
We therefore continuously monitor the calibration source spectrum after the launch and measure the CTI increase for each observation.
The CTI function we use for Xtend takes into account both transfer-time and area dependencies \citep{Kanemaru2020}.
The slope of the Mn K$\alpha$ line centroid, measured from its time history, was used to estimate the CTI increase rate for each CCD, while the CI was also being performed during the measurement.
The result is $(3$--$5)\times10^{-6}$~yr$^{-1}$ for each calibration region, consistent with predictions from ground-based radiation damage experiments \citep{Kanemaru2019}.
It has been demonstrated that the SXI CCDs with the notch structure (explained in Section~\ref{sec:sxi}) are radiation tolerant in orbit and will meet the mission requirements until the EOL of the XRISM mission.
Figure~\ref{fig:55Fe_trend} shows the long-term trend of the Mn K$\alpha$ line centroid and its FWHM for the calibration source regions on each CCD.
The results indicate that, after applying standard pipeline processes such as CTI and charge trail corrections, these parameters remain stable within 1--2~ch over a year (even after changing the CI rows as noted in Section~\ref{sec:demo}). 
In other words, the energy resolution and gain of Xtend demonstrate long-term stability.

By applying a response matrix prepared for a standard analysis and  fitting the Mn K$\alpha$ spectra with a single Gaussian model, we determined the in-orbit energy resolution to be $\sim170$--180~eV (FWHM), which is significantly better than the mission requirement of 200~eV at the satellite's beginning of life (BOL).
We summarize the energy resolution of Xtend, measured both in orbit and during the ground tests in Table~\ref{tab:55Fe}.
As noted by \citet{Nakajima2018} and \citet{Tanaka2018}, the SXI CCDs onboard Hitomi exhibited a degraded energy resolution of approximately 10~eV compared with their ground tests.
This is presumably because the Hitomi/SXI data suffered from the light leakage, and for XRISM/Xtend, addressing this issue as noted above prevents the  degradation.

\begin{figure}[t]
 \begin{center}
  \includegraphics[angle=0,width=8.2cm]{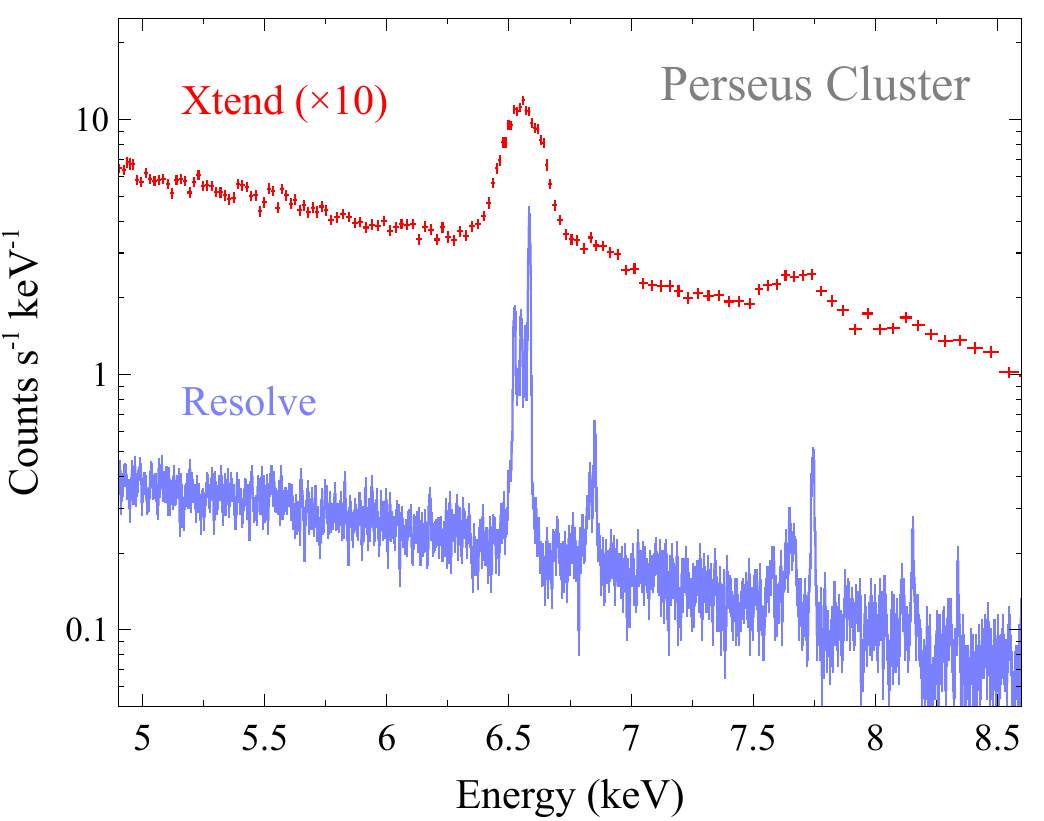} 
 \end{center}
\caption{Example X-ray spectra around ionized Fe (and Ni) K lines of the center region of the Perseus Cluster obtained with Xtend (red) and Resolve (light blue). The count rate for Xtend  is multiplied by a factor of ten for display purpose.
{Alt text: X-ray spectra of the Perseus Cluster.} 
}\label{fig:perseus}
\end{figure}

To check and keep monitoring the energy resolution and the effect of the CTI correction outside the calibration source regions, we use extended celestial sources, primarily clusters of galaxies and nearby SNRs.
An example spectrum is shown in figure~\ref{fig:perseus}, focusing the energy band around the  ionized Fe lines of the Perseus Cluster (OBSID: 000154000), extracted from the on-axis region.
The result indicates that, compared to the Resolve spectrum, the Fe He$\alpha$ and Ly$\alpha$ lines are resolved (more clearly seen in the case of Abell~2319; Figure\ref{fig:abell2319_spec}), and the line complex around Fe He$\beta$ \citep[cf.][]{Hitomi2016} is also prominent.
After applying the latest standard pipeline processes,  we measured  the energy resolution at the on-axis position  to be approximately 170~eV@6.4~keV, which is consistent with those obtained in the calibration source regions (Table~\ref{tab:55Fe}).
The result is also consistent for the other off-axis regions, and thus we conclude that there is no significant spatial difference in the spectroscopic performance of Xtend across the FoV.

\subsection{Effective Area}

While the  ground measurements for the mirror Xtend XMA indicate that its effective area is $\sim600$~cm$^{2}$@1.5~keV and $\sim400$~cm$^{2}$@6.4~keV \citep{Boissay-Malaquin2024}, the quantum efficiency of the CCDs should be taken into account.
The Xtend CCDs have a thick depletion layer of $\sim200~\mu$m, which enables a high X-ray stopping power up to $\sim13$~keV, whereas the CBF (polyimide$+$aluminum) and the OBL (aluminum) are positioned along the light path of the SXI.
The  resultant total effective area of Xtend was expected to be $\sim420$~cm$^{2}$@1.5~keV and $\sim310$~cm$^{2}$@6.0~keV.
These measurements derived from ground calibration tests separately, and the response function was prepared before the launch \citep{XRISM2022}.
We thus verified the total effective area of Xtend as the combination of the mirror (XMA) and detector (SXI) using in-orbit data from celestial objects.

\begin{figure}[t]
 \begin{center}
  \includegraphics[angle=0,width=8.8cm]{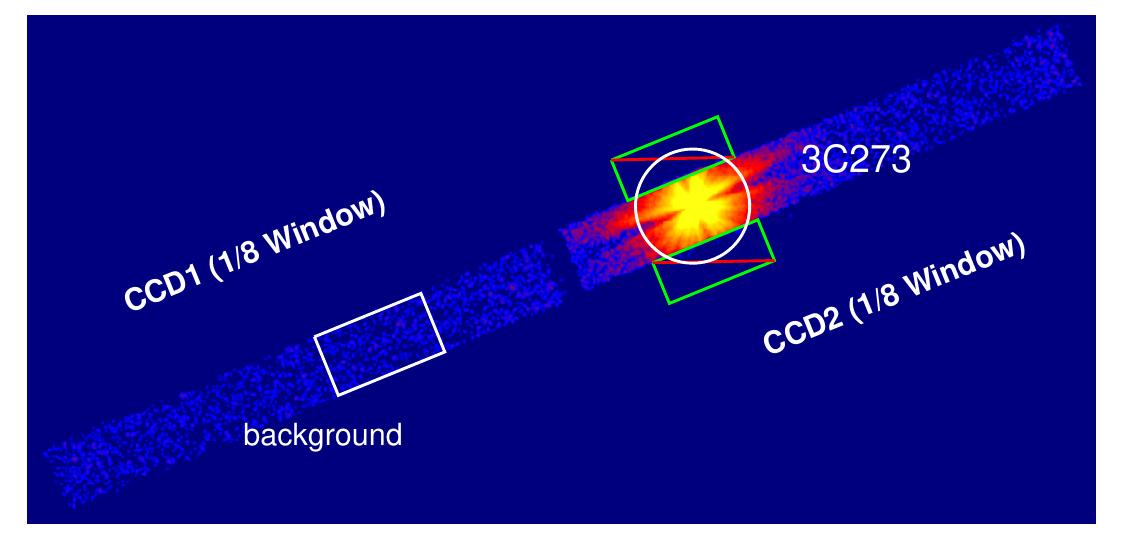} 
 \end{center}
\caption{Xtend image of a point source quasar, 3C273, obtained in the "1/8 Window No Burst" mode, displaying only CCD1 and CCD2. The spectral and background extraction regions are enclosed by the white circle and white rectangle, respectively. The regions enclosed by the green lines are excluded from our analysis since they are outside the window.
{Alt text: X-ray image of 3C273.} 
}\label{fig:3c273_image}
\end{figure}

In order to estimate the on-axis effective area of Xtend in orbit, we performed a joint observation of a bright quasar, 3C273, with XRISM (ObsID: 000145000) and other X-ray satellites: Swift (ObsID: 00089771001), Chandra (ObsID: 29166), XMM-Newton (ObsID: 0810822101), and NuSTAR (ObsID: 11002608002) on January 7--8, 2024.
Instruments we used for this calibration are Swift X-Ray Telescope \citep[XRT;][]{Burrows2005}, Chandra High Energy Transmission Grating \citep[HETG;][]{Canizares2005}, XMM-Newton European Photon Imaging Camera \citep[EPIC MOS and PN;][respectively]{Turner2001, Struder2001}, and NuSTAR Focal Plane Modules \citep[FPMA/B][]{Harrison2013}.
Since previous studies of this quasar suggest a variable photon index ($\Gamma\sim1.3$--1.6) correlated with its luminosity \citep[][and references therein]{Page2004}, we first confirmed that the temporal variation is below 7\%, which does not affect our analysis. 
Figure~\ref{fig:3c273_image} shows the Xtend image of 3C273 obtained in the ``1/8 Window No Burst'' mode during the observation campaign  above.
A circular region with a radius of $2\arcmin$ was selected for the source spectrum of 3C273, excluding the regions outside of the window, while the background was extracted from a distant region on the adjacent CCD1 chip.

\begin{figure*}[t]
 \begin{center}
    \includegraphics[angle=0,width=15cm]{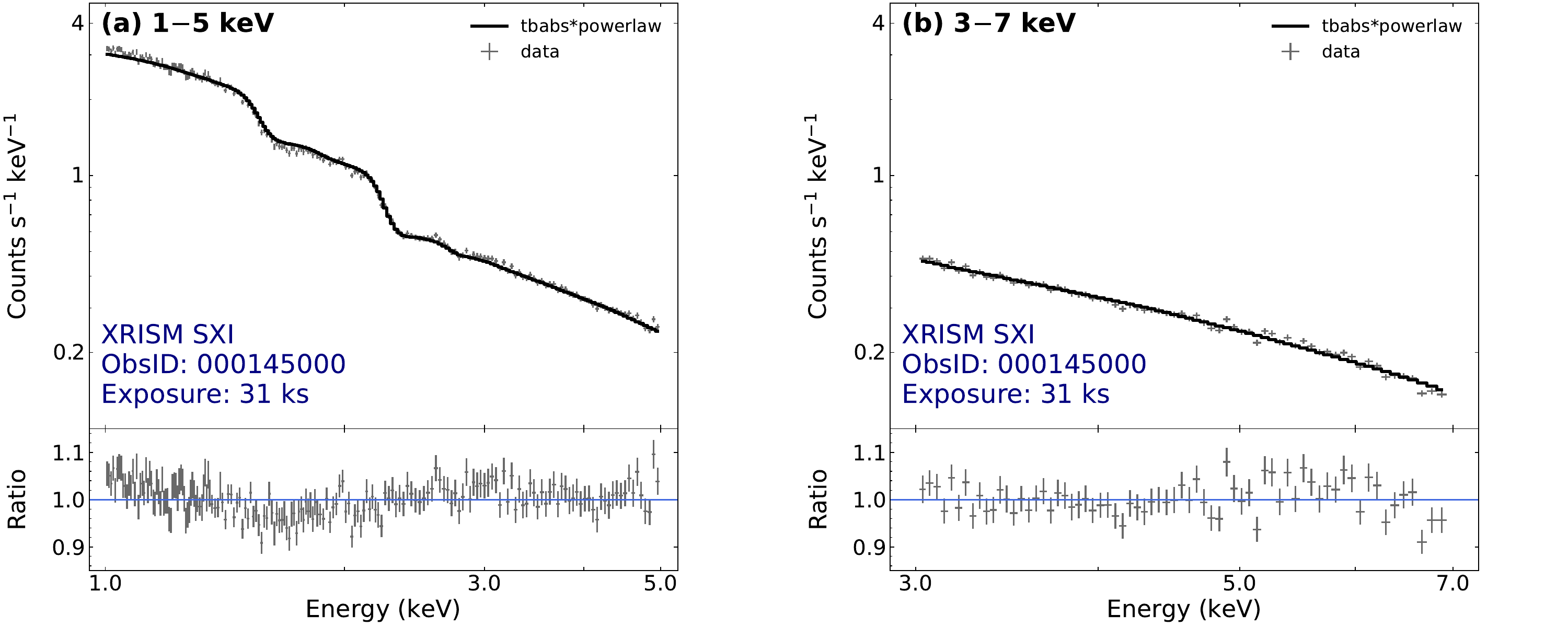} 
 \end{center}
\caption{Xtend spectra of 3C273 fitted with an absorbed power-law model, represented by the black lines. The results of the soft (1--5~keV) and hard (3--7~keV) energy bands are shown on the left (a) and right (b), respectively. Residuals are also displayed in the lower panel of each figure.
{Alt text: X-ray spectra of 3C273.} 
}\label{fig:3c273_spec}
\end{figure*}

\begin{figure*}[t]
 \begin{center}
  \includegraphics[angle=0,width=15cm]{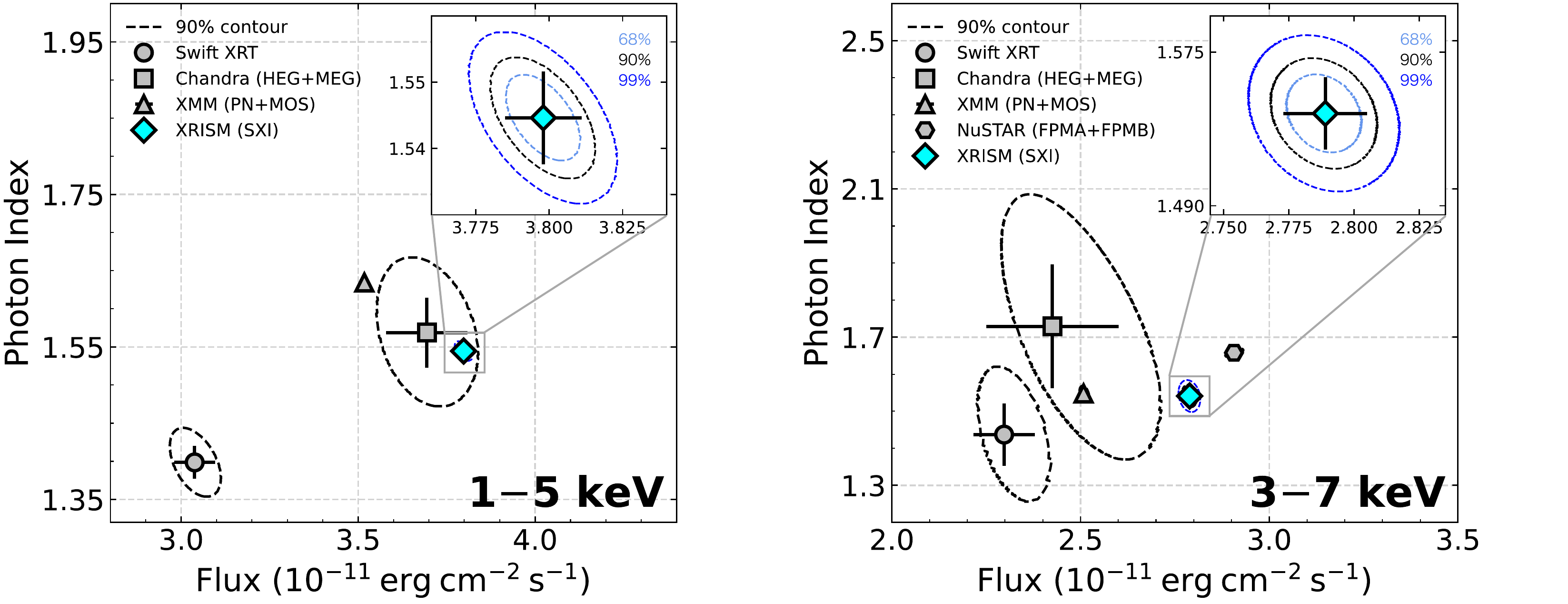} 
 \end{center}
\caption{Top: Scatter plots of the best-fit photon indices $\Gamma$ versus X-ray fluxes  for all instruments used for the joint observations of 3C273. The left and right panels are the results in the soft (1--5~keV) and hard (3--7~keV) energy bands, respectively. The broken black lines represent the 90\% contours for each detector. Those for XMM-Newton and NuSTAR are smaller than the markers and thus not visible.
{Alt text: Graphs depicting a comparison of best-fit parameters among  instruments.} 
}\label{fig:3c273_gamma}
\end{figure*}

To evaluate the on-axis effective area of Xtend, we separately fitted the soft band (1--5~keV) and hard band (3--7~keV) Xtend spectra using an absorbed \citep[T\"{u}bingen-Boulder absorption, \texttt{tbabs};][]{Wilms2000} single power-law model and compared the best-fit parameters with those obtained from the other instruments (figure~\ref{fig:3c273_spec}).
As summarized in figure~\ref{fig:3c273_gamma}, the results show that using an effective area file (i.e., ancillary response file; ARF) generated by standard XRISM analysis tools yields results similar to those from other satellites. 
We confirm that both $\Gamma$ and X-ray fluxes are consistent within $\sim20\%$, in line with previously-reported systematic differences between X-ray satellites in cross-calibration observations of a pulsar wind nebula G21.5$-$0.9 \citep{Tsujimoto2011}.
These results indicate that the on-axis effective area of Xtend is consistent with that expected before the launch.
Note that in the hard X-ray band, we further validated this result using another target, PKS~2155$-$304 (ObsID: 000127000), with the same method.
A detailed analysis, including the relative effective area uncertainty below 1~keV for both Xtend and Resolve, will be summarized in a separate paper (Miller et al., \textit{in prep.}).



\subsection{Quantum Efficiency}
\begin{figure*}[t]
 \begin{center}
  \includegraphics[angle=0,width=14cm]{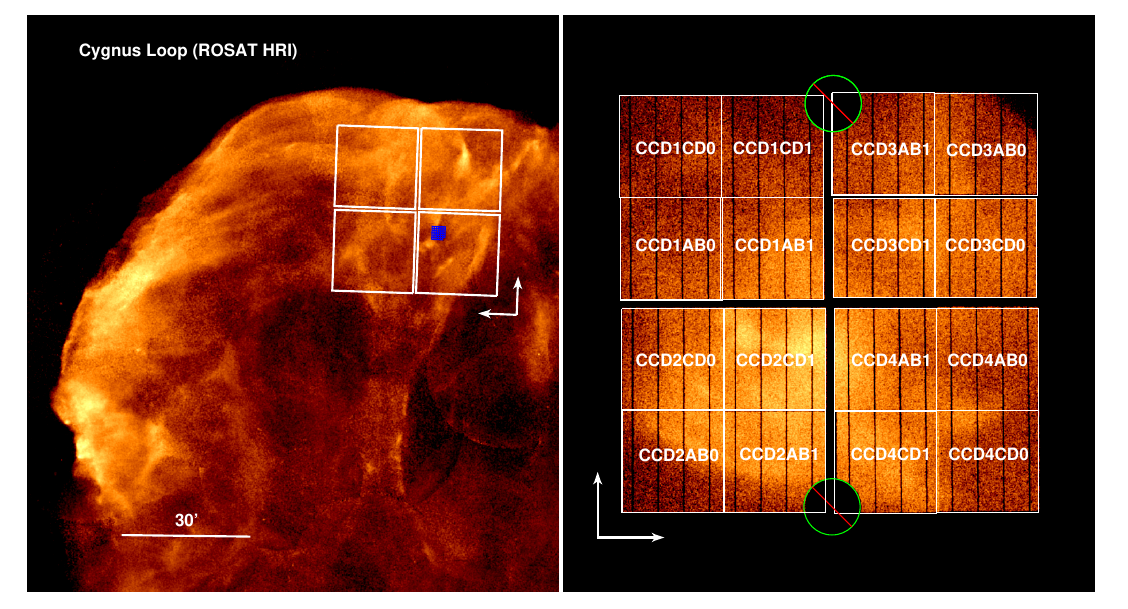} 
 \end{center}
\caption{Left: Soft X-ray image of the Cygnus Loop obtained with ROSAT HRI in the sky coordinate system. The Xtend and Resolve FoVs are overlaid with the white and blue squares, respectively. Right: Xtend image of the observed region in the detector coordinate system. 
Note that rotating the image 90$\degree$ counterclockwise approximately aligns the ROSAT HRI image.
The calibration source regions, shown by the green circles, were excluded from our analysis. The 16 sectors used for calibration are separated by the white lines, and each is labeled. 
{Alt text: X-ray image of the Cygnus Loop.} 
}\label{fig:cyg_image}
\end{figure*}

\begin{figure*}[t]
 \begin{center}
  \includegraphics[angle=0,width=9.5cm]{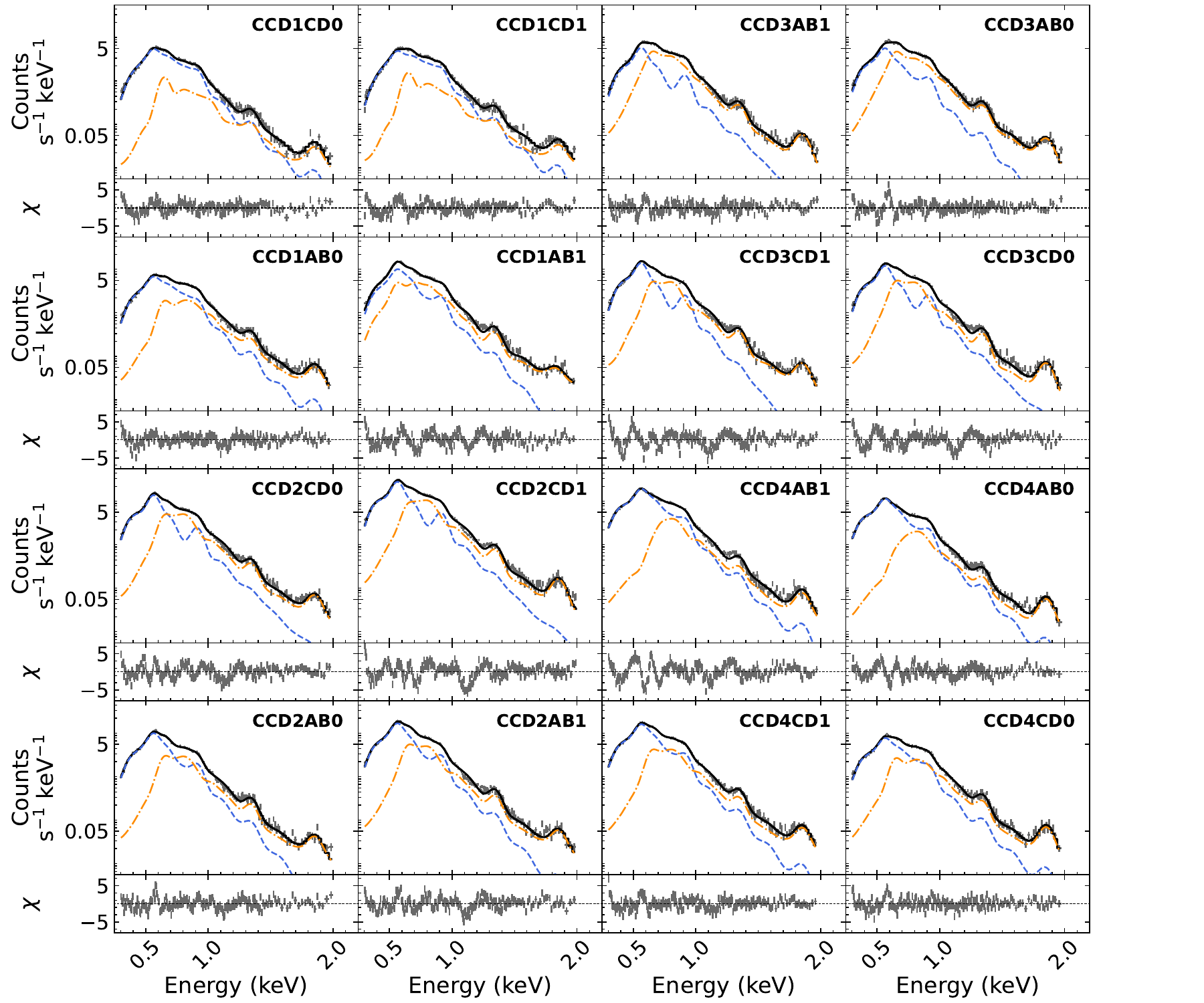} 
 \includegraphics[angle=0,width=8cm]{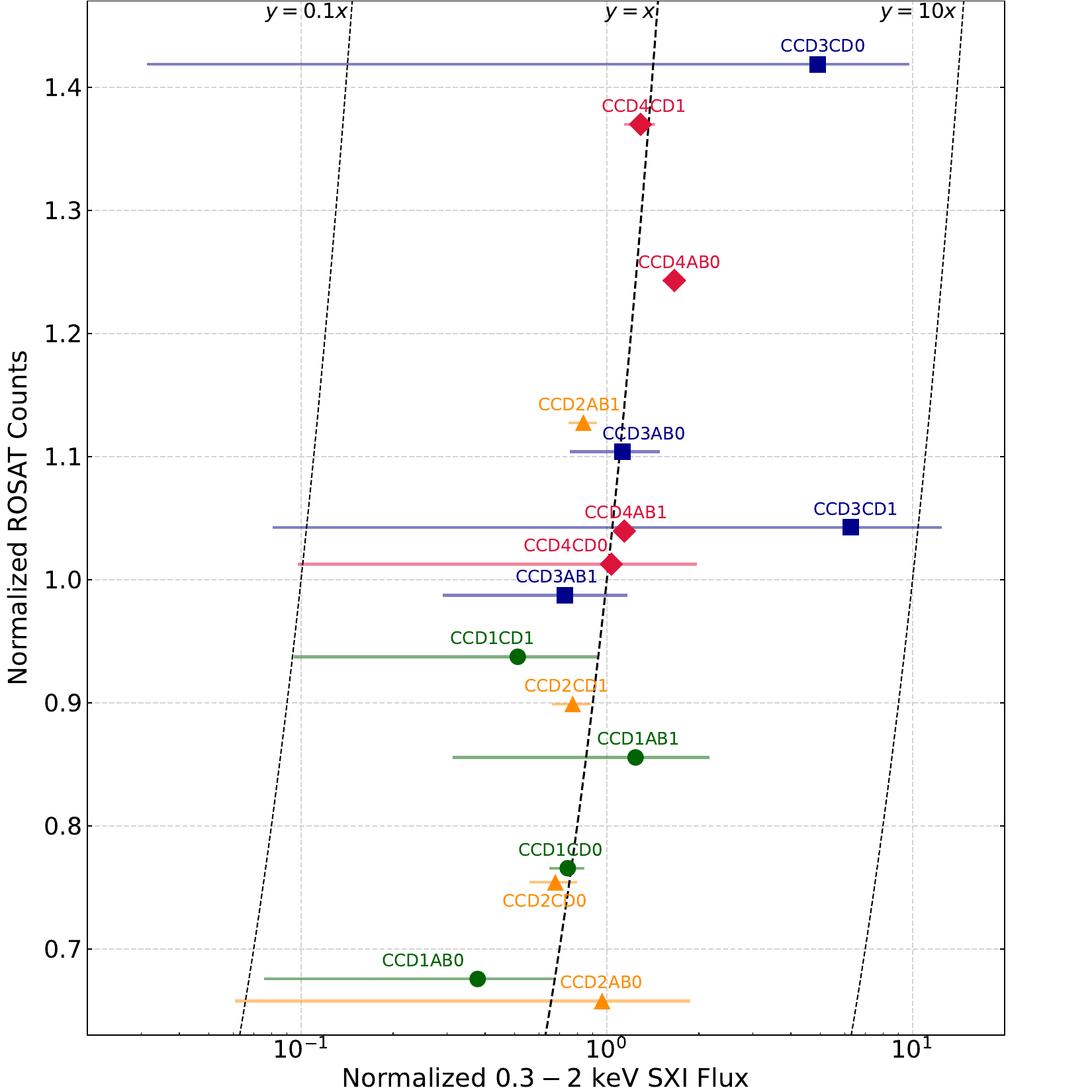} 
 \end{center}
\caption{Left: Xtend spectra of the 16 sectors (see figure~\ref{fig:cyg_image}) in the Cygnus Loop fitted with an absorbed two-temperature \texttt{nei} model. The best-fit models are overlaid in the black lines. The cold and hot components are represented by the blue dashed and orange dash-dotted lines, respectively. Residuals are also displayed in the lower panel of each figure. Right: Comparison between the normalized count rates of the Cygnus Loop at different positions, measured with ROSAT, and the X-ray fluxes obtained from the spectral fits. The green circle, orange triangles, blue squares, and red diamonds represent the results from CCD1, CCD2, CCD3, and CCD4, respectively. We note that the uncertainties of the ROSAT data are negligible. The dotted lines represent a constant ratio of normalized ROSAT counts to normalized Xtend flux, where the notation ``y = 0.1x'' denote that the normalized X-ray flux measured with Xtend is 10\% of the expected value with ROSAT.
{Alt text: X-ray spectra of the Cygnus Loop and  graph showing obtained fluxes.} 
}\label{fig:cyg_spec}
\end{figure*}

To investigate the uniformity of quantum efficiency across different positions on each CCD chip of Xtend, we observed a nearby bright extended SNR, the Cygnus Loop (ObsID: 100008010), and compared the X-ray fluxes with a count map obtained with the High Resolution Imager (HRI) on the ROentgen SATellite \citep[ROSAT;][]{Truemper1982}.
As displayed in the left panel of figure~\ref{fig:cyg_image}, the Xtend FoV covers a relatively flat and bright region in the Cygnus Loop, where the complex filamentary X-ray morphology of this remnant is clearly visualized using Xtend.
We divided the FoV into 16 sectors, as shown in the right panel of  figure~\ref{fig:cyg_image}, and performed spectral fits to estimate the X-ray flux at each position.
The left panel of figure~\ref{fig:cyg_spec} presents the best-fit results using an absorbed two-temperature non-equilibrium ionization model  (\texttt{nei}), following the previous analysis shown by \citet{Uchida2009}.

\begin{figure}[t]
 \begin{center}
  \includegraphics[angle=0,width=7.5cm]{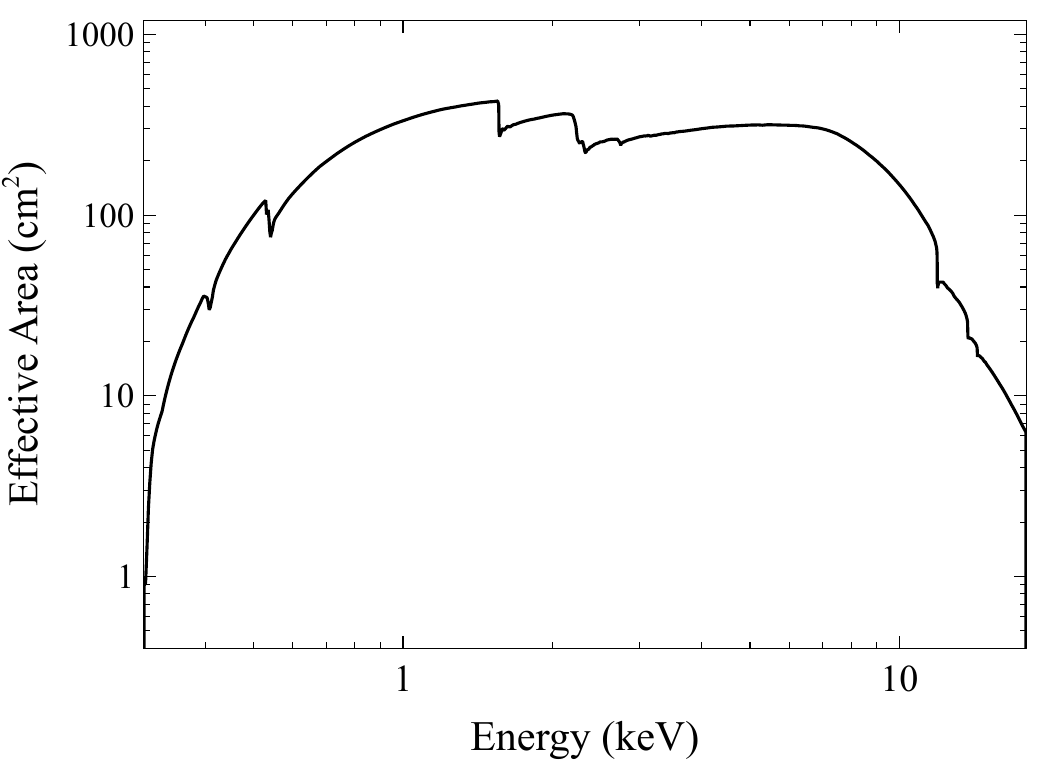} 
 \end{center}
\caption{Effective area of Xtend in the energy range of 0.4--13~keV. The ARF calculation assumes an on-axis point source and considers the entire Xtend FoV.
{Alt text: One line graph.} 
}\label{fig:arf}
\end{figure}

In order to estimate the uniformity of Xtend's quantum efficiency, we calculated the flux for each region divided by the median of all the results. 
We then compared this normalized flux with the HRI count rate in the same sector normalized by their median.
As a result, we confirmed that the normalized X-ray fluxes are consistent with the expected values, and the 3$\sigma$ ranges of most sectors (14 out of 16; 88\%) intersect with the line y=x (the right panel of figure~\ref{fig:cyg_spec}).
Since the spectra of the Cygnus Loop dominate in the soft X-ray band (0.3--2.0~keV), we also conducted a similar analysis on the Perseus Cluster (2.0--7.0~keV) and found that the results are consistent even in the hard X-ray band.
These results clearly indicate that the effective area in orbit is in good agreement with that obtained in the ground tests, showing no significant differences across different positions in the FoV.
We show an example plot of the total effective area of Xtend, taking into account the mirror response and the quantum efficiency of SXI, in figure~\ref{fig:arf}.


\subsection{Contamination}
Since X-ray CCDs on satellites are generally operated at low temperatures, typically below $-100~{\rm\degree C}$, the accumulation of outgas contamination onto their surfaces over the course of the mission becomes a significant issue, as it reduces the quantum efficiency of the detector, particularly in the soft energy band below 1~keV \citep[e.g.,][]{Plucinsky2018}.
Although the SXI-S has the CBF, which prevent the intrusion of outgassing into the camera body, the Xtend CCDs keep at a constant low temperature during the mission period, making them more prone to contamination material adhering.
We therefore need to continuously monitor the effects of contamination using a well-observed celestial calibration source.
For this purpose, we use the brightest  SNR in the Small Magellanic Cloud, 1E~0102.2$-$7219 to estimate the thickness of the contamination layer on axis.

\begin{figure}[t]
 \begin{center}
     \includegraphics[angle=0,width=7.5cm]{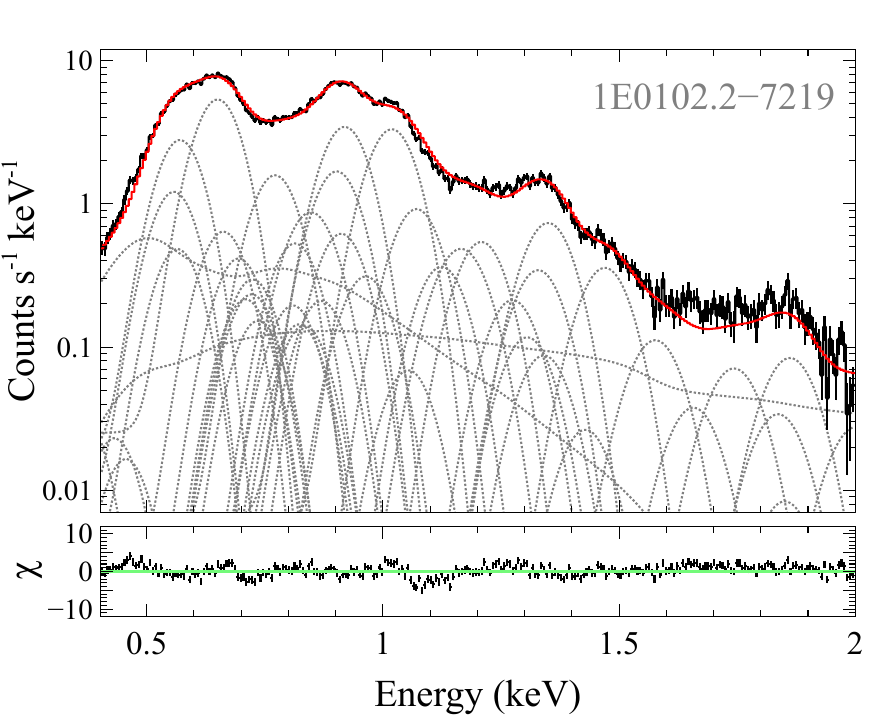}  
     \includegraphics[angle=0,width=7.5cm]{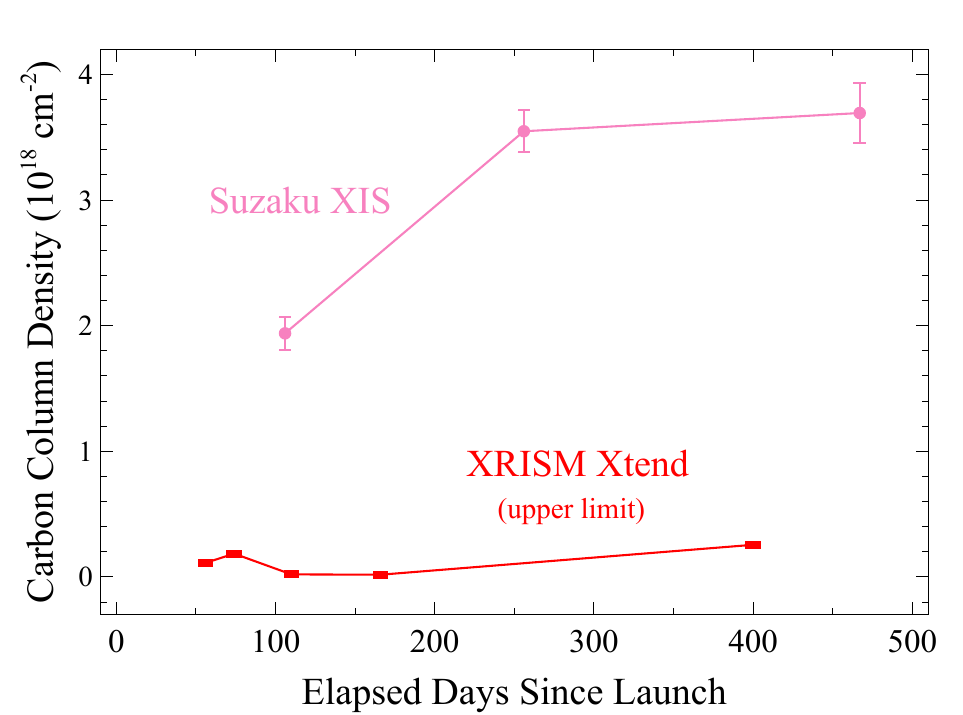}  
 \end{center}
\caption{Top: Xtend spectrum of  SNR 1E~0102.2$-$7219. The solid red line illustrates the best-fit IACHEC model, while the grey lines represent its components. The residuals are given in the lower panel. Bottom: Temporal variation of on-axis contamination levels (column density of carbon) of the CCDs onboard Suzaku (magenta) and XRISM (red) after their launches. Note that all values for XRISM Xtend represent   upper limits for each observation.
{Alt text: X-ray spectrum and graph indicating a temporal variation of the contamination.} 
}\label{fig:contami}
\end{figure}

SNR 1E~0102.2$-$7219 has been observed several times with XRISM since its launch; we extracted soft-band (0.4--2.0~keV) spectra and fitted them with a standard  model \citep[cf.][]{Plucinsky2017} provided by the International Astronomical Consortium for High Energy Calibration (IACHEC).
An example spectrum and the best-fit model are shown in figure~\ref{fig:contami}.
Since a previous experiment for the Suzaku satellite \citep{Mitsuda2007} suggests that the most plausible contaminants in a spacecraft are high-molecular compounds containing oily substances \citep{Urayama2013}, we assume them to be carbon-rich materials.
Under this assumption, we varied the column density of carbon in the IACHEC model and estimated the thickness of the contamination layer.
The result indicates that no significant contamination is observed on-axis, even in the most recent observation data obtained on October 11, 2024, which is consistent with our expectation before the launch. 
As displayed in figure~\ref{fig:contami}, the obtained upper limit of the carbon column density is $\lesssim 2 \times 10^{17}$~cm$^{-2}$, which is at least one order of magnitude lower than the values obtained with Suzaku for the same time interval since launch.

\begin{figure}[t]
 \begin{center}
  \includegraphics[angle=0,width=8cm]{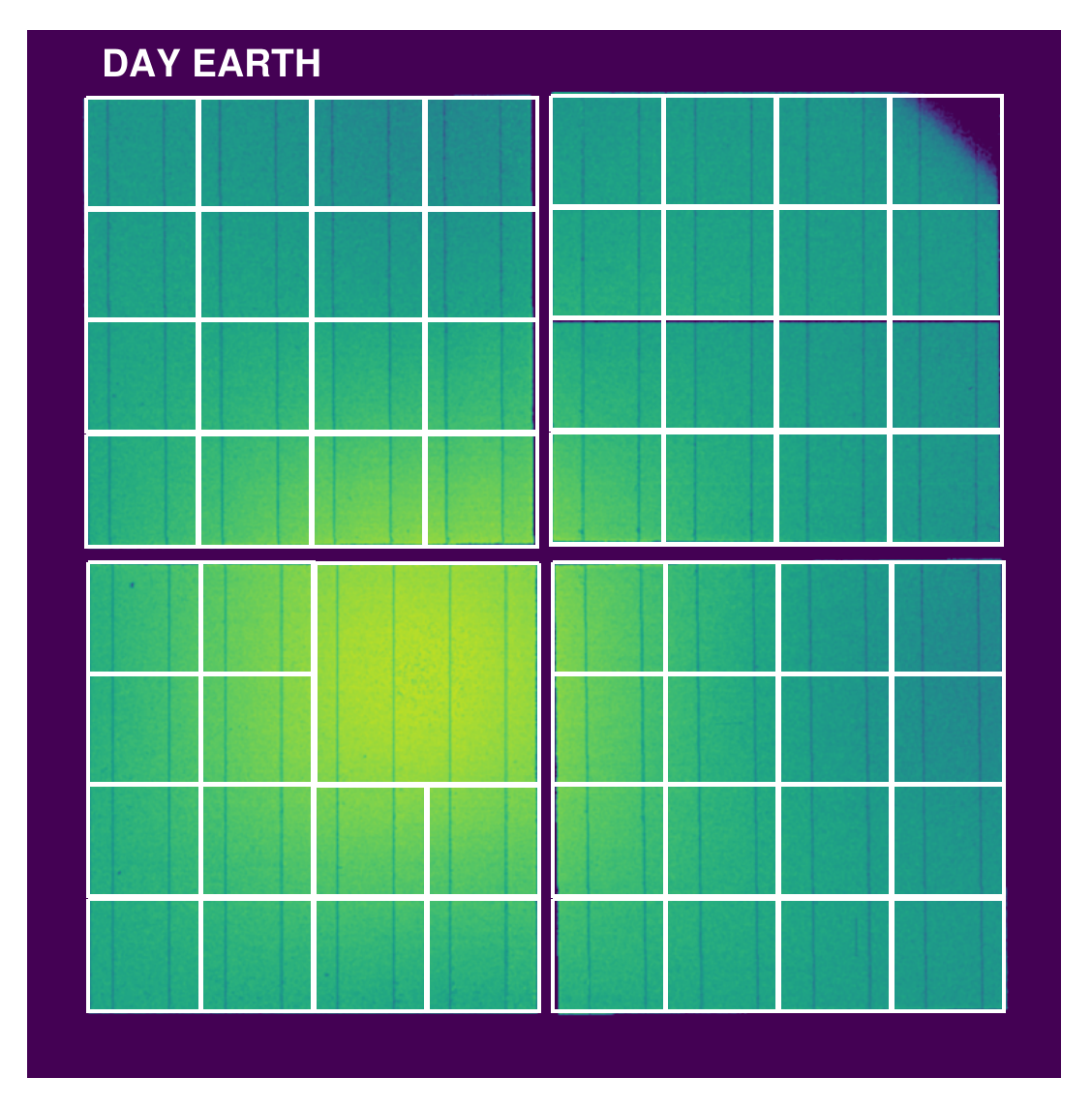}
     \includegraphics[angle=0,width=7.5cm]{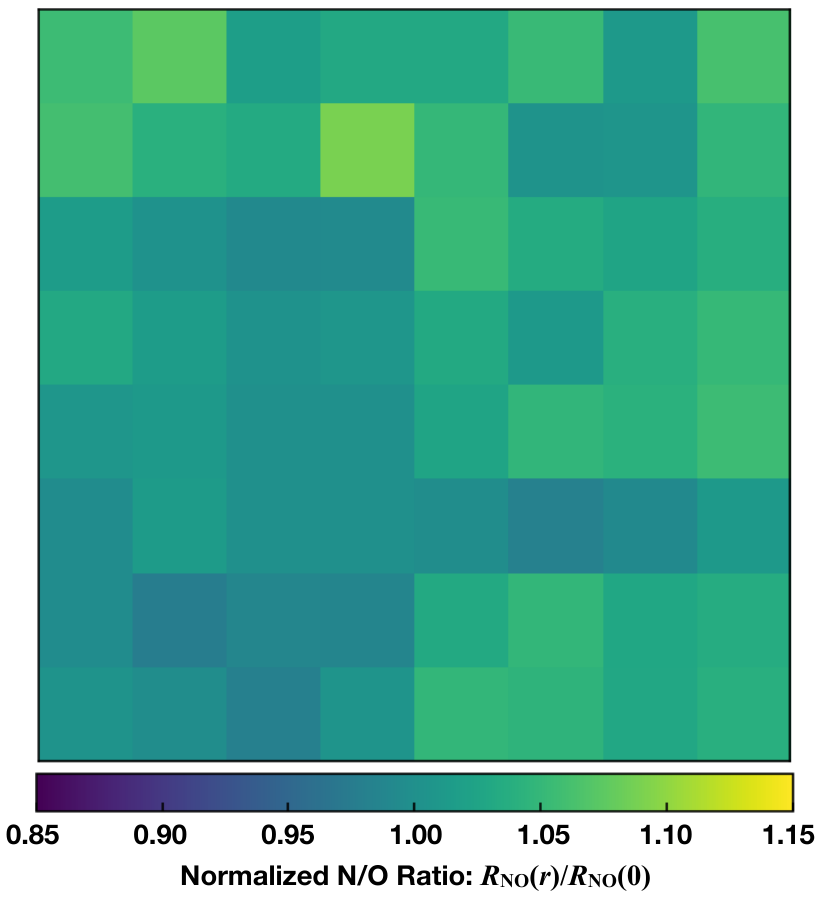}  
 \end{center}
\caption{Top: Xtend image obtained when the spacecraft was pointing toward the day Earth (0.4--2.0~keV). The image is displayed in the detector coordinate system. Regions used to estimate position-dependent contamination are enclosed by the white lines.  Bottom: Measured N/O ratios, $R_{\rm NO}(r)$, normalized by $R_{\rm NO}(0)$ determined at the on-axis nominal position (see equation~\ref{eq:contami}).  
{Alt text: X-ray image and color map.} 
}\label{fig:dayearth}
\end{figure}

We also investigated the effect of contamination across the entire FoV of Xtend by analyzing soft-band images obtained during the day Earth approximately one year after the launch of XRISM, as shown in figure~\ref{fig:dayearth}.
Since the day-Earth X-ray spectra are typically dominated by fluorescence lines of nitrogen (N; 0.39~keV) and oxygen (O; 0.53~keV), the ratio of the count rates of each line, N/O, is a good indicator for estimating the relative amount of contamination compared to that on axis.
Here, we assume that the contaminant is diethylhexyl phthalate (DEHP; C$_{24}$H$_{38}$O$_{4}$), a plasticizer commonly used in polymers and identified as a major contributor to contamination on the CCD of Hinode \citep{Urayama2010, Kosugi2007}.
If an observed N/O ratio in a region is denoted as $R_{\rm NO}(\bm{r})$, where $\bm{r}$ is a position vector from the nominal point $\bm{r}=\bm{o}$, the difference in contamination thickness, $d(\bm{o})-d(\bm{r})$ can be described as follows:
\begin{eqnarray}
d(\bm{o})-d(\bm{r}) &=& \frac{{\rm ln}\left(R_{\rm NO}(\bm{r})/R_{\rm NO}(\bm{o})\right)}{\mu_{\rm N}-\mu_{\rm O}},\label{eq:contami}
\end{eqnarray}
where $\mu_{\rm N}$ and $\mu_{\rm O}$ are the linear attenuation coefficients of DEHP at the energies of the N and O fluorescence lines, respectively.
The obtained $R_{\rm NO}(\bm{r})$ values at each position were normalized to the value at the nominal position. 
The Xtend FoV was divided into 61 sectors, and the resulting normalized N/O ratios, $R_{\rm NO}(\bm{r})/R_{\rm NO}(\bm{o})$, are presented as a map in the lower panel of figure~\ref{fig:dayearth}, showing a spatial variation is lower than 10\% and the difference in contamination thickness is lower than $5\times10^{19}$~cm$^{-2}$ at each position.
Consequently, we conclude that the contamination is negligible at any location within the FoV even after one year since the launch.



\subsection{Non X-ray Background}\label{sec:nxb}
\begin{figure}[t]
 \begin{center}
  \includegraphics[angle=0,width=8.5cm]{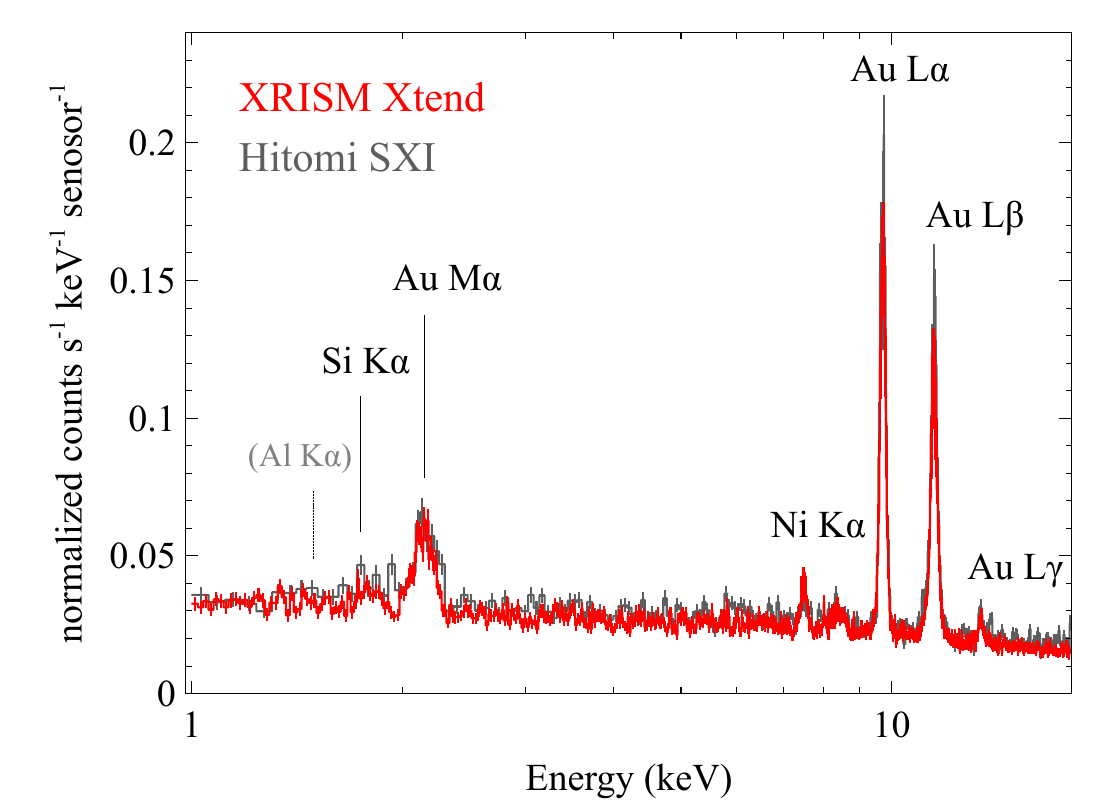} 
 \end{center}
\caption{NXB spectra obtained with XRISM/Xtend (red) and Hitomi/SXI (grey). The data were extracted from the entire FoV, excluding the calibration source regions. The count rates are normalized per sensor, for which we assume a $1280\times1280$ pixel detector with a physical pixel size of $24\mu{\rm m}\times24\mu{\rm m}$. The vertical lines indicate the literature values for the centroid energies of each elemental fluorescence, while Al K$\alpha$ is not significant.
{Alt text: X-ray spectra of NXB.} 
}\label{fig:nxb}
\end{figure}

Since XRISM operates in a low Earth orbit (LEO), where the spacecraft follows a circular trajectory at an altitude of  $\sim550$~km, it is expected to achieve a low and stable background (i.e., NXB), as demonstrated previously with Suzaku, which operated in a similar orbit \citep{Tawa2008}.
We have developed the mechanical design specifications of the Xtend SXI to mitigate events caused by cosmic rays or fluorescent X-rays originating from detector materials excited by particles.
The thickness of the SXI camera body was thus designed to be $>10$~g~cm$^{-2}$, based on a Monte Carlo simulation of the  interaction of cosmic/X-rays with a CCD by \citet{Anada2008}.
To remove NXB events, we have also introduced a method generally known as the ``ASCA grade'' selection technique, which eliminates multi-pixel events caused by cosmic rays and selects bona fide X-ray events.
As a result, a low NXB level is achieved with XRISM/Xtend, which is nearly identical to that of Hitomi/SXI as shown in figure~\ref{fig:nxb}.
While the intensity of NXB  generally depends on the cut-off rigidity (COR), as extensively investigated for the XIS \citep{Tawa2008}, we confirmed that the spectral shape of the Xtend NXB is unaffected by the COR (less than 10\%).
Note that the screening criteria for extracting an NXB spectrum is defined as 
\begin{eqnarray}
({\rm SAA\_SXI == 0)}~\&\&~({\rm T\_SAA\_SXI >277)} \notag \\
\&\&~{\rm (ELV < -5)}~\&\&~{\rm (DYE\_ELV > 100)},
\end{eqnarray}
within the extended housekeeping data distributed to users.\footnote{A brief explanation of these  criteria can be found in Table~6.3 of the Hitomi Data Reduction Guide (https://heasarc.gsfc.nasa.gov/docs/hitomi/analysis/).}

As seen in figure~\ref{fig:nxb}, several emission lines are observed in the Xtend NXB spectrum.
The most prominent lines, Au L$\alpha$ and Au L$\beta$, primarily originate from the gold-coated inner surface of the SXI-S camera body, which is made of stainless steel and coated to shield the background radiation.
The cold plate, on which the CCDs are mounted, is coated with gold via vapor deposition, which also potentially contributes to the Au fluorescence lines (see also figure~\ref{fig:abell2319_spec}).
The relatively weak lines of Au M$\alpha$ and Au L$\gamma$ are also observed at 2.2~keV and 11.5~keV, respectively.
The Ni K$\alpha$ line at 7.5~keV also originates from the housing materials, where a nickel coating is applied to the stainless steel body beneath the gold coating (known as a graded-Z shield).
The weak Si K$\alpha$ line is also observed, which are produced by fluorescence within the CCD wafers.
Although aluminum is used for both the CBF and the OBL, no emission lines are detected in the NXB spectrum obtained from a single observation with a 100-ks exposure.

\begin{figure*}[t]
 \begin{center}
  \includegraphics[angle=0,width=15cm]{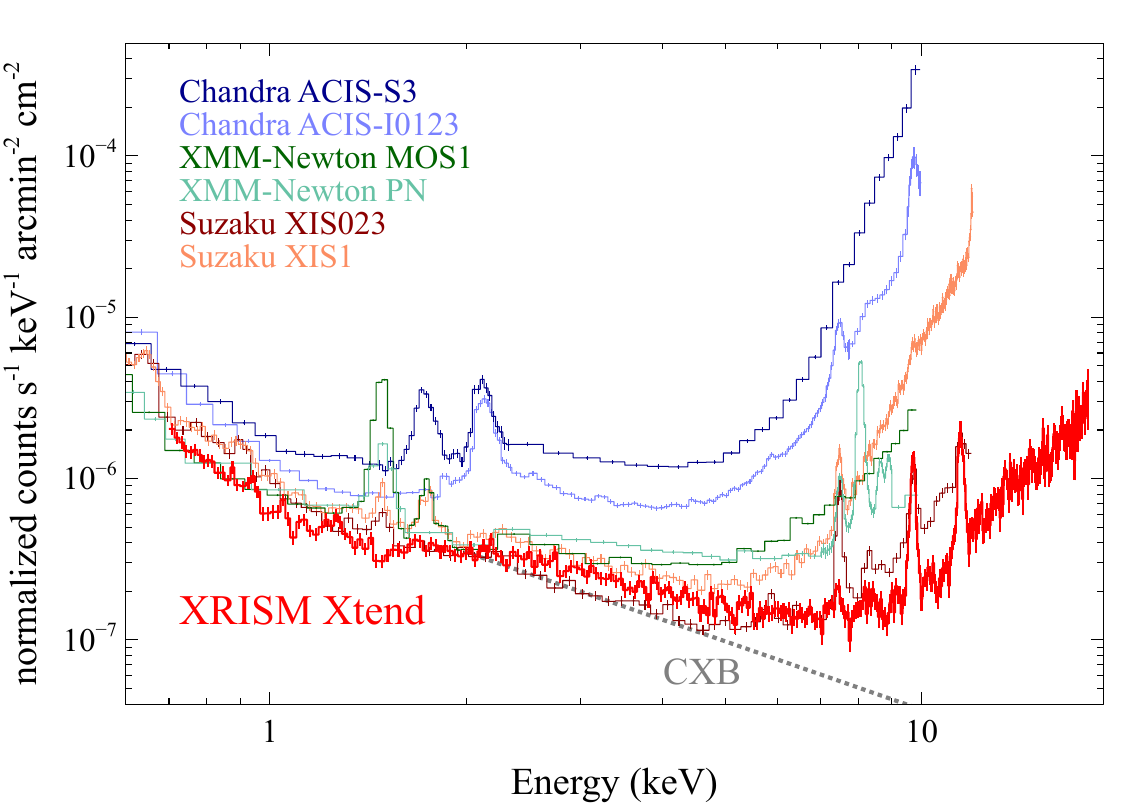} 
 \end{center}
\caption{Sky background spectra normalized by the effective area and the solid angle of the FoV. The data were estimated from an off-center region in the observation of 3C273. The dotted grey line represents the CXB level, and data obtained with other X-ray CCD instruments in orbit are also plotted for comparison.
 {Alt text: Graphs comparing sky background of each X-ray CCD instrument. }  
}\label{fig:all_nxb}
\end{figure*}

Figure~\ref{fig:all_nxb} compares the sky background (NXB$+$CXB) spectra obtained with major X-ray instruments onboard previous and current satellites: Chandra/ACIS-I \& S3, XMM-Newton/EPIC-MOS \& PN, Suzaku/XIS-FI (XIS0, XIS2, XIS3) \& BI (XIS1), and XRISM/Xtend.
These data, except for those of XRISM/Xtend, are derived from \citet{Katayama2004}, \citet{Tawa2008} and \citet{Nakajima2018}.
In this plot, we also display the CXB level of $9.3\times10^{-7}$~(E/1~keV)$^{-1.4}$~counts~s$^{-1}$~keV$^{-1}$~arcmin$^{-2}$~cm$^{-2}$ as a reference.
Since the spectra are normalized by the effective area of each instrument, the plot represents the surface brightness of the sky background, thereby reflecting the sensitivity of each detector to diffuse targets.
As a result, the background level for Xtend in the hard X-ray band (5.0--10~keV) has been determined to be ($1.52\pm0.03) \times 10^{-7}$~counts~s$^{-1}$~keV$^{-1}$~arcmin$^{-2}$~cm$^{-2}$, which are significantly lower compared to satellites in highly elliptical orbits (HEOs), such as Chandra and XMM-Newton.
The Xtend background level is also lower compared to other BI CCDs, like Suzaku XIS1, especially  above 6~keV, and is consistent with or rather lower than that  of the font-illuminated (FI) CCD, Suzaku/XIS-FI.
On the other hand, in the lower energy band below 6~keV, Suzaku/XIS-FI exhibits a lower background level, this is likely due to its narrower FoV compared to that of Xtend: a larger FoV results in higher photon counts from the CXB per unit effective area.
It is also notable that the fluorescence lines in the NXB spectrum of Xtend are fewer and weaker compared to the other instruments, which would be advantageous for spectroscopy.
We thus conclude that XRISM/Xtend combines a large effective area with a low background across a wide energy range.

\section{Summary and Future Prospects}
\subsection{Summary of the In-orbit Performance of Xtend}
We successfully activated  Xtend  (XMA$+$SXI) onboard XRISM in orbit and began observing celestial objects along with Resolve since 2023, after overcoming the setback of the Hitomi mission in 2016.
After the four CCDs (CCD1--4; CCD\_ID~0--3) reached the required operational temperatures, we conducted first light observations of the cluster of galaxies Abell~2319 and the SNR N132D, demonstrating the imaging and spectroscopic capabilities of Xtend.
Although several issues were identified in the Hitomi/SXI data \citep{Tanaka2018, Nakajima2018}, such as the light leakage and crosstalk, we addressed them prior to the launch and confirmed that there was no significant impact on the Xtend data from the in-orbit calibration observations. 
The Xtend image of the extended diffuse source indicates a large FoV of $38.5\arcmin\times38.5\arcmin$ covering an energy range of $0.4$--$13$~keV.
Taking into account its large effective area, we also present  one of the largest grasps ${\rm S}\Omega$ of Xtend ($\sim60~{\rm cm^2~degree^2}$)  among the currently available focal plane X-ray detectors.
The angular resolution of Xtend is $1.4\arcmin$ HPD, which enables observers to study faint structures of diffuse sources and achieve better separation of celestial objects.

Based on the calibration observations, we obtained the energy resolution of Xtend for each CCD sensor to be $\sim170$--180~eV, which meets the XRISM mission requirement and is sufficient to separate ionized emission lines, such as the He-like Fe He$\alpha$ ($\sim6.7$~keV) and H-like Fe Ly$\alpha$ ($\sim6.9$~keV) lines.
We also tested the effective area of Xtend through a joint observation campaign of bright point sources, such as 3C273 and PKS~2155$-$304, with XRISM and other X-ray satellites (Swift, Chandra, XMM-Newton, and NuSTAR). 
Consequently, we confirmed that the effective area in orbit reaches $\sim420$~cm$^{2}$@1.5~keV and $\sim310$~cm$^{2}$@6.0~keV.
We also conclude that  there was no significant outgas contamination onto the surface of the SXI CCDs even one year after the launch.
The count rate of Xtend's NXB is nearly identical to that of Hitomi/SXI, and a low background level has been achieved. 
These results indicate that Xtend is suitable for detecting faint and diffuse X-ray sources.

\subsection{Future Prospects as  a Wide Field X-ray Imager}
In summary, Xtend has great potential for observing faint diffuse sources, thanks to its large grasp (figure~\ref{fig:grasp}), sufficient angular resolution (figures~\ref{fig:abell2319_img_vig} and \ref{fig:N132D}), moderately good energy/spatial resolutions (figures~\ref{fig:abell2319_spec} and \ref{fig:N132D_color}), and low background level (figures~\ref{fig:nxb} and \ref{fig:all_nxb}).
These aspects have already been shown in a previous study, for instance, \citet{Suzuki2025}, in which they report a detection of faint diffuse X-ray emission around the potential PeVatron microquasar V4641~Sagittarii \citep[e.g.,][]{Alfaro2024, LHAASO2024}.
Another potential observation using Xtend is introduced by \citet{Zhou2024}, who claim that an Xtend observation combined with Resolve is crucial for probing decaying dark matter, including sterile neutrinos and axion-like particles, with a focus on the large effective area and FoV of Xtend.
As shown by the first-light XRISM observations of Abell~2319 (figures~\ref{fig:abell2319_img_vig}) and N132D \citep{Audard2024}, the better angular resolution of Xtend can be not only useful in supporting the Resolve analysis but also allows Xtend to produce significant scientific results on its own.

The largest grasp of Xtend shown in figure~\ref{fig:grasp} is also valuable for future follow-up observations and transient searches in the context of  time-domain and multi-messenger astrophysics, which is expected to become increasingly important in the coming years.
As introduced by \citet{Tsuboi2024}, Xtend has discovered over several dozen X-ray transients so far using the XRISM/Xtend Transient Search (XTS) system, with reports published in the Astronomer's Telegram (ATel): the first one is of an X-ray flare from a plausible optical counterpart, LP~593$-$2, an M3.9 dwarf binary \citep{Yoshimoto2024}.
As these outcomes suggest, the performance of Xtend will prove valuable for observing interesting transient celestial objects in the future, for instance, X-ray follow-up observations of supernovae in nearby galaxies \citep[e.g.,][]{Zimmermann1994, Grefenstette2023} and high-energy neutrino emitters as astrophysical sources of ultrahigh-energy cosmic rays  \citep{Yoshida2024}.
We therefore expect that Xtend will be widely used in future time-domain and multi-messenger astrophysics and will \textit{extend} the scope of observational capabilities, opening new windows for unprecedented discoveries.

\begin{ack}
The authors thank Junko Hiraga, Yuki Amano, Hiromichi Okon, Takuto Narita, Kai Matsunaga, Yujiro Saito, Moe Anazawa, Hirotake Tsukamoto, Jin Sato, Toshiyuki Takaki, Yuta Terada, Honoka Kiyama, Kaito Fukuda, Mariko Saito, Yamato Ito, Mariko Saito, Shuusuke Fudemoto, Satomi Onishi, Junichi Iwagaki, Koki Okazaki, Kazunori Asakura, Maho Hanaoka, Yuichi Ode, Mio Aoyagi, Tomohiro Hakamata, Shunta Nakatake, Toshiki Doi, Kaito Fujisawa, and Mitsuki Hayashida for their contribution to the development of the XRISM/Xtend/SXI. 
We also thank Eric D. Miller, Kenji Hamaguchi, Naomi Ota and Kyoko Matsushita for their valuable inputs.
We thank Yang Soong, Hideyuki Mori, Larry Olsen, Richard Koenecke, Wilson Lara, Leor Bleier, Marshall Sutton, Marton Sharpe, Larry Lozipone, Sean Fitzsimmons, Tony Baltusis, Dan Dizon, Gary Sneiderman, Meng Chiao, Steve Kenyon, Danielle Gurgew for their support throughout the XMA development and ground calibration at NASA's Goddard Space Flight Center. 
We thank Iurii Babyk, Steven J. Kenyon, Edward W. Magee, Connor Martz, Brian R. McNamara, Hideyuki Mori, Nicholas E. Thomas, and Makoto Sawada for support on the XMA thermal shield measurements taken at Lawrence Berkeley National Laboratory's Advanced Light Source (ALS) and at the Canadian Light Source (CLS), and gratefully acknowledge support scientists at these facilities, including Eric Gullikson (ALS Beamline 6.2.3), Teak Boyko, David Muir, Ronny Sutarto, Feizhou He, and Jeff Warner (CLS Resonant Elastic and Inelastic X-ray Scattering beamline). 
Part of this work was performed under the auspices of the U.S. Department of Energy by Lawrence Livermore National Laboratory under Contract DE-AC52-07NA27344.
This research used resources of the Advanced Light Source, which is a DOE Office of Science User Facility under contract no. DE-AC02-05CH11231.
Part of the research described in this paper benefitted from measurements at the Canadian Light Source, a national research facility of the University of Saskatchewan, which is supported by the Canada Foundation for Innovation (CFI), the Natural Sciences and Engineering Research Council (NSERC), the Canadian Institutes of Health Research (CIHR), the Government of Saskatchewan, and the University of Saskatchewan.

\end{ack}

\section*{Funding}
This work is supported by Japan Society for the Promotion of Science (JSPS) KAKENHI with the Grant number of 19K03915, 23K22536, 24K21547 (H.U.), 19K21884, 20H01947, 20KK0071, 23K20239, 24K00672 (H.N.), 23K20850, 21H01095 (K.M.), 20KK0071, 24H00253 (H.N.), 21J00031, 22KJ3059, 24K17093 (H.S.), 21K13963, 24K00638 (K.H.), 21K03615, 24K00677 (M.N.), 20H00175, 23H00128 (H.M.), 21H04493, 15H02090, 14079204 (T.G.T.), 22H01269 (T.K.), 24K17105 (Y.K.), 24KJ1483 (S.I.). K. K. N. acknowledges the support by the Yamada Science Foundation.


%
%
%
%
%



\bibliographystyle{apj}
\bibliography{sample631}


\end{document}